\documentclass[a4paper,preprintnumbers,showpacs,onecolumn,superscriptaddress,nofootinbib,amsmath,amssymb,notitlepage]{revtex4-1}

\usepackage{enumitem}
\usepackage{times}
\usepackage{dcolumn}
\usepackage{mathrsfs}
\usepackage{comment}
\usepackage{hyperref}
\usepackage{amsmath,accents}
\usepackage{mathtools}
\usepackage{bbm}
\usepackage{bm}
\usepackage{amsfonts}
\usepackage{amssymb}
\usepackage{mathrsfs}
\usepackage{wasysym}
\usepackage{graphicx}
\usepackage[]{subfigure}
\usepackage{filecontents}
\usepackage[dvipsnames]{xcolor}
\usepackage{bbold}

\hypersetup{
    bookmarks=true,         
    pdftoolbar=true,        
    pdfmenubar=true,        
    pdffitwindow=true,     
    pdfstartview={FitH},     
    pdftitle={My title},     
    pdfauthor={author},      
    pdfsubject={Subject},    
    pdfcreator={Creator},    
    pdfproducer={Producer},  
    pdfkeywords={keyword1} 
    pdfnewwindow=true,       
    colorlinks=true ,        
    linkcolor=blue  ,        
    citecolor=blue,      
    filecolor=green,       
    urlcolor=blue            
}

\newcommand{\arccosh}{\operatorname{arccosh}}

\newcommand{\arctanh}{\mathrm{arctanh}}

\newcommand{\ed}{\mathrm{d}}

\newcommand{\ve}{V_\mathrm{eff}}

\newcommand{\We}{Weierstra$\ss$}
\newcommand{\im}{\mathrm{i}}
\newcommand{\IS}{\mathrm{IS}}
\newcommand{\OS}{\mathrm{OS}}

\begin{document}

\title{Study of null and time-like geodesics in the exterior of a Schwarzschild black hole with quintessence and cloud of strings}

\author{Mohsen Fathi}
\email{mohsen.fathi@postgrado.uv.cl}
\affiliation{Instituto de F\'{i}sica y Astronom\'{i}a, Universidad de Valpara\'{i}so,
Avenida Gran Breta\~{n}a 1111, Valpara\'{i}so, Chile}

\author{Marco Olivares}
\email{marco.olivaresr@mail.udp.cl}
\affiliation{Facultad de
              Ingenier\'{i}a y Ciencias, Universidad Diego Portales,  Avenida Ej\'{e}rcito
              Libertador 441, Santiago, Chile}

\author{J.R. Villanueva}
\email{jose.villanueva@uv.cl}
\affiliation{Instituto de F\'{i}sica y Astronom\'{i}a, Universidad de Valpara\'{i}so,
Avenida Gran Breta\~{n}a 1111, Valpara\'{i}so, Chile}


\begin{abstract}
Recently, an analytical study of radial and circular orbits for null and time-like geodesics that propagate in the spacetime produced by a Schwarzschild black hole associated with cloud of strings, in a universe filled by quintessence, has been done in Ref.~\cite{Mustafa:2021}. In this paper, we complete the aforementioned study by investigating possible analytical solutions to the equations of motion for other types of bound orbits, beside taking into account the cases of unbound orbits. This requires an extensive study of the corresponding effective potentials that categorize the test particle motion. We follow the standard Lagrangian dynamics to parametrize the radial and angular geodesics and the resultant (hyper-)elliptic integrals of motion are treated accordingly. We also simulate the orbits which correspond to different levels of energy in the effective potentials.

\bigskip

{\noindent{\textit{keywords}}: Black holes, quintessence, cloud of strings, particle geodesics}\\

\end{abstract}

\maketitle

\section{Introduction and Motivation}\label{sec:intro}

The general theory of relativity got famous soon after its advent, firstly because of its bewildering formalism with its strong roots in the Riemannian geometry, and secondly, because of its unprecedented success in predicting rather small deflection angles and precession in the perihelion, respectively in the lensing process around the Sun and in the planetary orbit of Mercury. And the theory still continues with its triumphs in conformity with observational results regarding its prediction of gravitational waves \cite{Abbott:2016blz,Abbott:2016nmj,Abbott:2017vtc} and black hole dynamics \cite{Akiyama:2019} (see Ref.~\cite{Bluem:2020} for a recent review). The two first brilliant predictions of general relativity are, however, of fundamental importance, since they result directly from the Riemannian nature of the theory which incarnates in the geodesic equations. In this sense, the study of particle trajectories in the spacetimes predicted by general relativity has been of great interest among theoretical scientists, as well as observational astrophysicists. In particular, the attention to the derivation of analytical solutions to the equations of motion in general relativistic geometries, uplifted significantly after the calculations done by Hagihara in 1930 for particle trajectories in Schwarzschild spacetime \cite{Hagihara:1930}.
Since then, efforts have been made rigorously to find analytical solutions for test particle geodesics in strong gravity. These include the significant formulation of lens equations, particle scatterings and bound orbits around black holes \cite{Bardeen:1972a,Bardeen:1973a,Bardeen:1973b,Chandrasekhar:2002}. On the other hand, the more new general relativistic solutions were proposed, the more it became complicated treating, analytically, the particle geodesics. Although the advancements in computational equipments, softwares and methods, have paved the way in doing numerical simulations of particle orbits even in stationary spacetimes, the analytical treatments of these phenomena are still of considerable merit. Since such studies are based on firm mathematical foundations, they can serve to show the precise way the test particles behave. Accordingly, the Keplerian orbits of mass-less and massive particles (in the sense of null and time-like geodesics) in general relativistic black hole spacetimes, have been studied by means of advanced elliptic integration methods, for example, in Refs.~\cite{kraniotis_compact_2003,cov04,hackmann_complete_2008,hackmann_geodesic_2008,bisnovatyi-kogan_strong_2008,kagramanova_analytic_2010,hackmann_complete_2010,gibbons_application_2012,munoz_orbits_2014,de_falco_approximate_2016,chatterjee_analytic_2019} for static, and in Refs.~\cite{kraniotis_precise_2004,beckwith_extreme_2005,kraniotis_frame_2005,hackmann_analytical_2010,kraniotis_precise_2011,enolski_inversion_2011,kraniotis_gravitational_2014,barlow_asymptotically_2017,uniyal_null_2018,hsiao_equatorial_2020,gralla_null_2020,kraniotis_gravitational_2021,fathi_analytical_2021} for stationary spacetimes. 

One important feature that appears in most of the aforementioned references, is the coupling of the black hole spacetime with at least one cosmological term. This feature, on its own, makes the analysis more intense and likely to encounter hyper-elliptic integrals. On the other hand, inclusion of cosmological-associated terms in general relativistic solutions have a long history, going back to the early years of the theory. And ever since the uplift in the attention to the dark side of the universe regarding the evidences like the flat galactic rotation curves, anti-lensing, and the accelerated expansion of the universe  \cite{Rubin1980,Massey2010,Bolejko2013,Riess:1998,Perlmutter:1999,Astier:2012}, relativists have become more inclined to the inclusion of cosmological components in the solutions to Einstein field equations, for them to be more compatible with astrophysical observations. In fact, the evolution of black holes are now seen to be affected by the dynamics of the universe where they reside \cite{JimenezMadrid:2005,Jamil:2009,Li:2019,Roy:2020}. From the the geometric point of view, this is approached by adding cosmological components to the black hole spacetime under consideration, to compensate for the dark features. This is commonly done by inclusion of a dark fluid energy-momentum tensor in the Einstein field equations, corresponding for example, to a dark matter halo \cite{Xu:2018,Das:2021}, or a quintessential field \cite{Kiselev:2003,Saadati:2019,AliKhan:2020}. This latter, in particular, has been postulated to explain the accelerated expansion of the universe, however, in a dynamic essence that is in contrast with the standard $\Lambda$-cold dark matter ($\Lambda$CDM) model, by replacing the cosmological constant with a quintessential scalar field. The search for alternatives to the cosmic fluid has been being developed constantly. As a special case, which is also of the interest of this paper, the cosmic fluid is supposed to be composed of a relativistic dust cloud consisting of one-dimensional strings (instead of point particles). This way, the Schwarzschild spacetime was generalized in Ref.~\cite{Stachel:1977}, to be associated with the so-called cloud of strings. The gauge-invariant generalization of this spacetime was proposed in Ref.~\cite{Letelier:1979}, and some of its geodesic phenomena were studied in Ref.~\cite{Batool:2017}.

In this work, we aim at the full study of null and time-like geodesic trajectories that are possible around a Schwarzschild black hole which is coupled with both the quintessential field and the cloud of strings. The exterior geometry of this black hole has been derived and discussed in Refs.~\cite{Toledo:2018,Dias:2019,Toledo:2019}. Furthermore, the radial and circular orbits of mass-less and massive particles have been studied analytically in Ref.~\cite{Mustafa:2021}. The shadow and the photon ring structure of this black hole has been investigated in Ref.~\cite{He:2021}. Indeed, this spacetime could be of interest, specially because of the fact that the quintessential component grants an extra potential to the solution, and hence, in addition to the accelerated expansion of the universe, it can compensate for the unseen galactic dark matter. Such property can be seen also in the Mannheim-Kazanas spherically symmetric solution to the fourth order Weyl conformal gravity, which was proposed to recover the flat galactic rotation curves \cite{Mannheim:1989}. In this spacetime, the source of gravity is supposed to be extended. Therefore, the black hole resides in a stringy universe. Recently, 
the respected parameters of this spacetime have been calibrated in the context of standard solar system observational tests \cite{Cardenas:2021}.

In this paper, in addition to the cases of radial and circular orbits studied in Ref.~\cite{Mustafa:2021}, we also present an analytical investigation of general angular orbits for both of the null and time-like geodesics, that include deflecting trajectories, critical and planetary orbits. Furthermore, for the aforementioned radial and circular orbits, we apply special elliptic integration methods that enable us expressing the solutions in the more compact and aesthetic \We{ian} forms. In order to complete the work done in the above reference, we go deeper into these kind of orbits by classifying them in accordance with their corresponding effective potentials. In particular, to provide more perception, the radial orbits are plotted for each of these cases.

To proceed with this investigation, in Sect. \ref{sec:BlackHole}, we briefly introduce the black hole spacetime and its causal structure. In Sect. \ref{sec:ParticleDynamics}, we construct the geodesic equations of motion in the context of the standard Lagrangian dynamics. We begin our study in Sect. \ref{sec:radial} with the radial geodesics for the both null and time-like cases. The corresponding effective potentials are categorized and the respected radial profiles of the temporal parameters are analyzed and plotted. In Sect.~\ref{sec:angular}, we continue our study by analyzing the effective potentials for particles with non-zero angular momentum. This way, several kinds of orbits become available which are studied analytically and in detail. Further  information are given in the appropriate places within the text. We summarize our results in Sect.~\ref{sec:conclusions}. We adopt the geometrized system of units in which, $G=c=1$.

\section{The black hole solution}\label{sec:BlackHole}

The static, spherically symmetric black hole solution associated with quintessence and cloud of strings, is described by the line element 
\begin{equation}\label{eq:metric}
	\ed s^2 = -B(r) \ed t^2 + \frac{\ed r^2}{B(r)} + r^2 \ed\theta^2 + r^2\sin^2\theta \ed\phi^2
\end{equation}
in the $x^\mu = (t,r,\theta,\phi)$ coordinates. The lapse function is given by \cite{Toledo:2018,Dias:2019}
\begin{equation}\label{eq:lapse}
	B(r) = 1-\alpha - \frac{2M}{r}-\frac{\gamma}{r^{3w_q+1}},
\end{equation}
with, $\alpha$,  $\gamma$ and $w_q$, representing the parameters of cloud of strings (dimensionless and $0<\alpha<1$), quintessence and the equation of state (EoS), and $M$ is the black hole's mass. {For a perfect fluid distribution of matter/energy and in the presence of quintessence, the EoS parameter respects the range $-1<w_q<-\frac{1}{3}$ \cite{Kiselev:2003}. The case of $\omega_q=-1$ corresponds to an extraordinary quintessence, or the cosmological constant. This particular case is similar to the Schwazschild-(anti-)de Sitter black hole, and its geodesic structure for mass-less and massive particles has been studied analytically in Refs.~\cite{kraniotis_compact_2003,hackmann_complete_2008}. When $\omega_q=-\frac{1}{3}$, the mathematical form of the lapse function becomes similar to a purely Schwarzschild black hole with cloud of strings, whose geodesic structure has been discussed in Ref.~\cite{Batool:2017}. Respecting the subject of the current study, both of the aforementioned cases propose all kinds of bound and unbound orbits. For the purpose of this paper, however, we confine ourselves to  the case of $w_q = -\frac{2}{3}$, that recovers
\begin{equation}\label{eq:lapse_1}
	B(r) = 1-\alpha - \frac{2M}{r}-\gamma r,
\end{equation}
and accordingly, $[\gamma]=\mathrm{m}^{-1}$. As it is observed, the term $\gamma r$ in the above lapse function, generates an additional gravitational potential. Such a potential is also available in the Mannheim-Kazanas static spherically symmetric solution to Weyl conformal gravity, and serves to compensate for the flat galactic rotation curves \cite{Mannheim:1989}. In this regard, the choice of $\omega_q=-\frac{2}{3}$ is of worth for further studies, besides the fact that it is relatively simpler than its other allowed values.} 

\begin{figure}[t]
	\begin{center}
		\includegraphics[width=8cm]{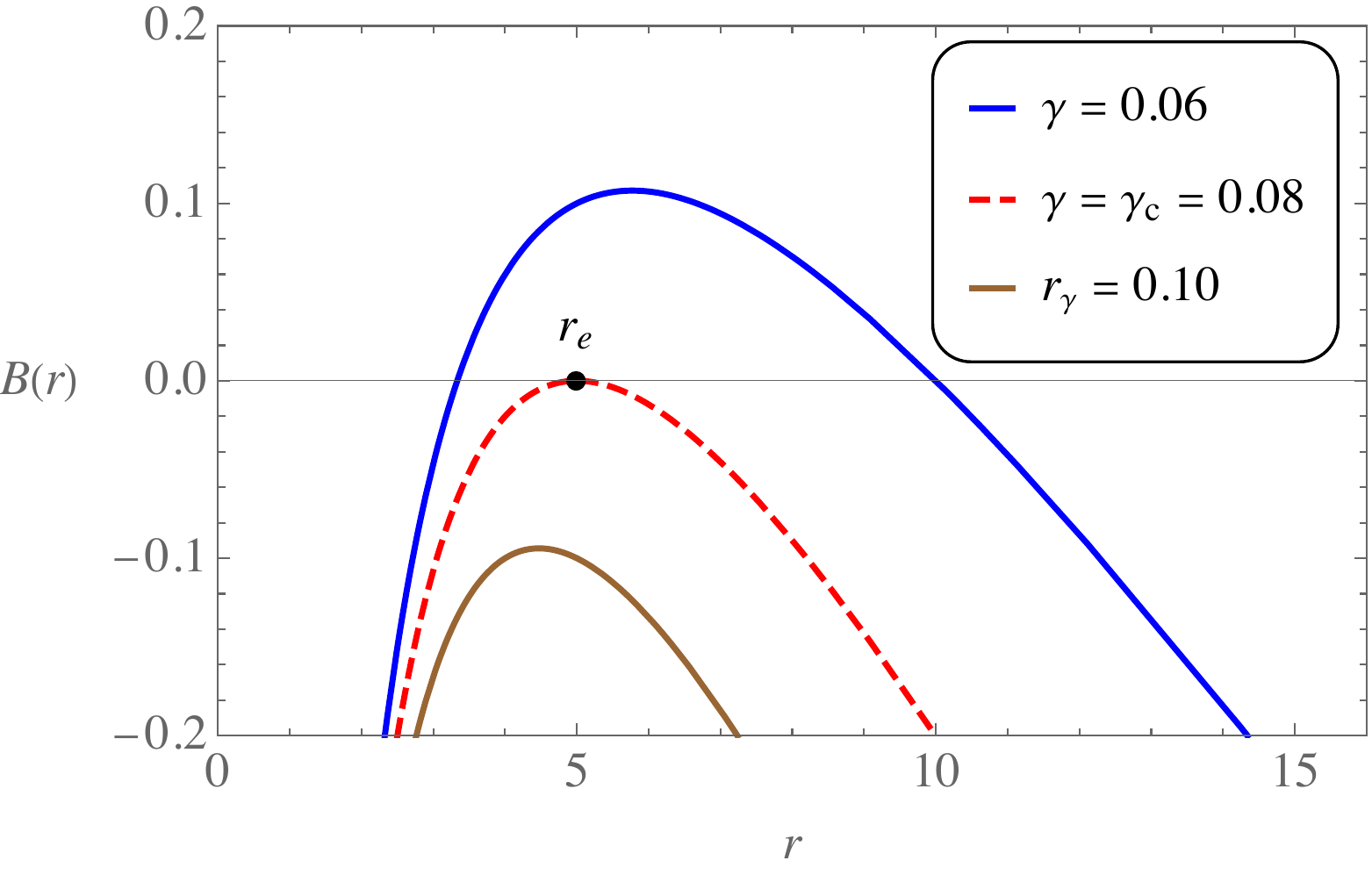}
	\end{center}
	\caption{Plot of the lapse function for different values of the $\gamma$  parameter. For this diagram and all the forthcoming ones, we have considered $\alpha = 0.2$, and the unit of length along the axes is $M$. This way, the extremal horizon will be located at $r_e=5$.}
	\label{fig:f3}
\end{figure}

The black hole horizons are located at the radial distances at which $B(r) = 0$, providing the two real roots \cite{Cardenas:2021}
\begin{eqnarray}
	&& r_{++} =\frac{1-\alpha}{\gamma}\cos^2\left(\frac{1}{2}\arcsin\left(\frac{2\sqrt{2M\gamma}}{1-\alpha}\right)\right),\label{eq:rg_sB}\\
	&& r_+ = \frac{1-\alpha}{\gamma}\sin^2\left(\frac{1}{2}\arcsin\left(\frac{2\sqrt{2M\gamma}}{1-\alpha}\right)\right),\label{eq:rH_sB}
\end{eqnarray}
that correspond, respectively, to the (quintessential) cosmological and the event horizons. This way, one can recast the lapse function \eqref{eq:lapse_1} as
\begin{equation}
B(r) = \frac{\gamma}{r}\left(r-r_+\right)\left(r_{++}-r\right).
    \label{eq:lapse_2}
\end{equation}
Note that, the extremal black hole for which $r_+=r_{++}=r_e = \frac{4M}{1-\alpha} $ corresponds to the case of $\gamma = \gamma_c=\frac{(1-\alpha)^2}{8M}$, and a naked singularity is obtained when $\gamma > \gamma_c$. Note that, for the cases of  $\alpha\rightarrow1$ and $\alpha\rightarrow0$, we get $\gamma_c\rightarrow0$ and $\gamma_c\rightarrow\frac{1}{8M}$. In Fig.~\ref{fig:f3}, the behavior of the lapse function \eqref{eq:lapse} has been plotted for different values of $\gamma$. The infinite redshift happens when the curves pass the $B(r)=0$ line for the first time (at the event horizon), whereas, the infinite blueshift happens when they pass this line for the second time (at the cosmological horizon). The corresponding extremal horizon $r_e$, has been also indicated in this diagram.

In the forthcoming sections, we analyze the analytical expressions for the radial and angular geodesics for both mass-less and massive particles around this black hole. For the particular case of radial null and time-like geodesics which have been also studied in Ref.~\cite{Mustafa:2021}, we present more compact analytical solutions in terms of \We{ian} elliptic functions, and we study different types of radial phenomena. Additionally, the bound and unbound angular orbits are given rigorous analytical studies regarding all kinds of the possible motions for mass-less and massive particles.

\section{Particle dynamics}\label{sec:ParticleDynamics}

In the static black hole spacetime \eqref{eq:metric}, the geodesic equations can be expressed in terms of the Lagrangian \cite{Chandrasekhar:2002}
\begin{eqnarray}\label{eq:Lagrangian}
    \mathcal{L} &=& \frac{1}{2} g_{\mu\nu} \dot{x}^\mu \dot{x}^\nu\nonumber\\
    &=& \frac{1}{2}\left[
    -B(r) \dot{t}^2 + \frac{\dot{r}^2}{B(r)} + r^2 \dot{\theta}^2 + r^2 \sin^2\theta\dot{\phi}^2
    \right]\nonumber\\
    &=& -\frac{1}{2}\epsilon,
\end{eqnarray}
where "dot" stands for differentiation with respect to the affine curve parameter $\tau$, which here is also regarded as the proper time. Furthermore, the null and time-like geodesics are characterized, respectively, by $\epsilon = 0$ and $\epsilon = 1$. Now, defining the generalized conjugate momenta 
\begin{equation}\label{eq:cte}
    \Pi_\mu = \frac{\partial\mathcal{L}}{\partial{\dot{x}}^\mu},
\end{equation}
one can introduce the two constants of motion 
\begin{subequations}\label{eq:cteMotion}
\begin{align}
    &  \Pi_t = -B(r) \dot{t} \equiv -E,\\
    &   \Pi_\phi = r^2\sin^2\theta\dot{\phi} \equiv L,
\end{align}
\end{subequations}
corresponding to the energy and angular momentum of the test particles. For the seek of convenience, we confine the geodesics to the equatorial plane, by fixing $\theta = \frac{\pi}{2}$, so that from Eqs.~\eqref{eq:Lagrangian} and \eqref{eq:cteMotion} we have
\begin{equation}\label{eq:rdot}
    \dot{r}^2 = E^2 - \ve(r),
\end{equation}
defining the gravitational effective potential
\begin{equation}\label{eq:Veff}
    \ve(r) = B(r)\left[\epsilon + \frac{L^2}{r^2}\right]. 
\end{equation}
Accordingly, one can obtain the radial and rotational equations of motion, as
\begin{eqnarray} 
	&&\left(\frac{{\rm d}r}{{\rm d} t}\right)^{2}= B^2(r)\left[1-\frac{\ve(r)}{E^2}\right],\label{rt}\\
	\label{rphi}
	&&\left(\frac{{\rm d}r}{{\rm d}\phi}\right)^{2}= \frac{r^4}{L^2}\left[E^2-\ve(r)\right].
\end{eqnarray}
Starting from the next section, these equations are treated in the context of the radial and angular motions, for both null and time-like trajectories.

\section{Radial motion}\label{sec:radial}

The test particles with $L=0$ are of great importance, since in the case of null geodesics they can construct the casual structure of the spacetime, and in the case of time-like geodesics, the difference between the perception of comoving and distant observers of infalling objects onto the black hole, is revealed. We begin with the null radial geodesics and continue with the time-like ones.

\subsection{Null geodesics}\label{subsec:radialNull}

\begin{figure}[t]
 \begin{center}
   \includegraphics[width=8cm]{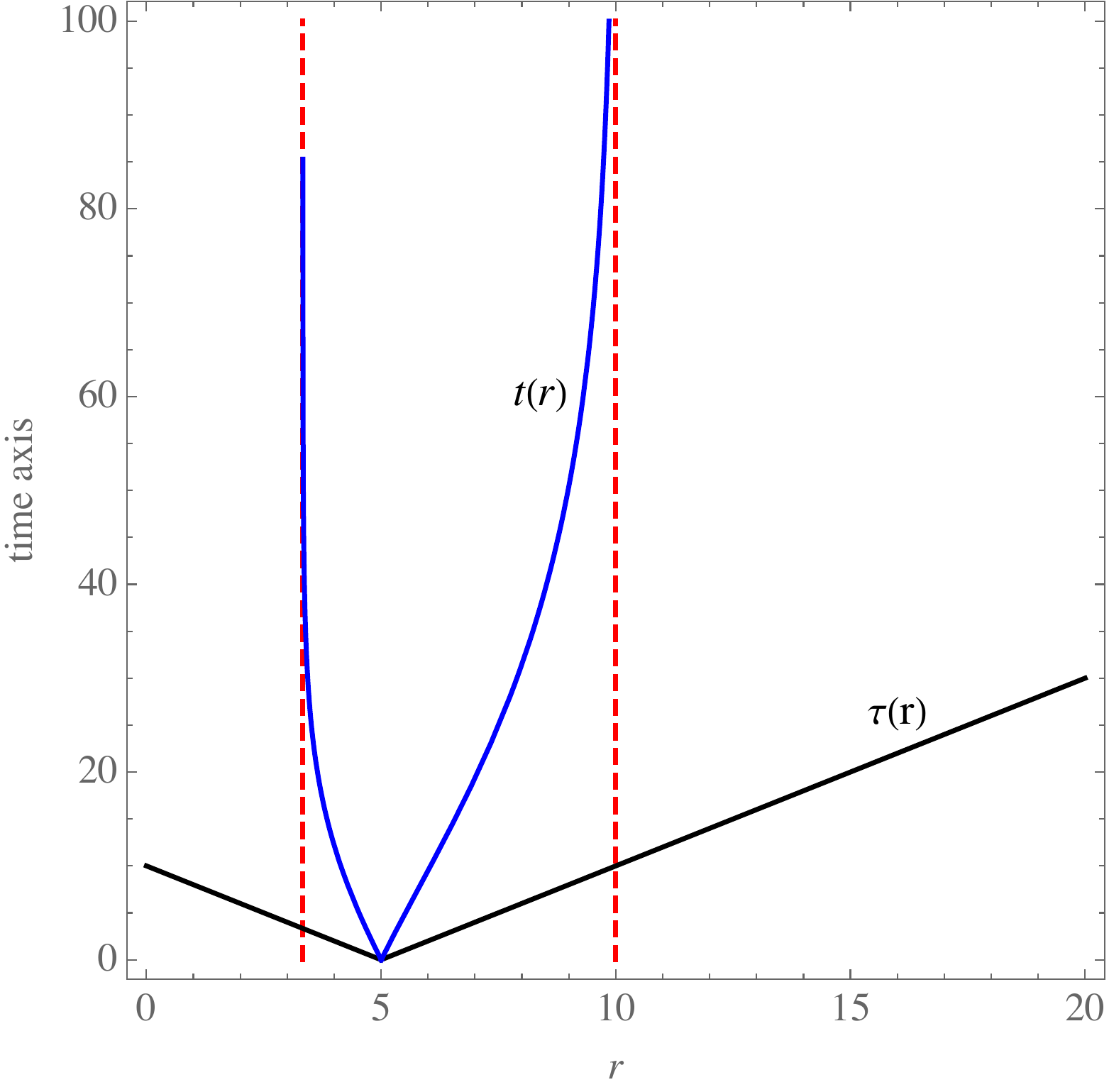}
 \end{center}
 \caption{The radial null geodesics plotted for $\gamma = 0.06$, $E=0.5$ and $r_i=5$. The diagrams indicate the asymptotic behavior of $t(r)$ (blue curves) and horizon crossing of $\tau(r)$ (black lines).  }
 \label{fig:time(r)-null}
\end{figure}

Letting $\epsilon = 0$, one can then see from Eq. \eqref{eq:Veff} that for the radially moving mass-less particles (e.g. photons), we have $\ve(r)=0$. Accordingly, Eqs. \eqref{eq:rdot} and \eqref{rt} become
\begin{eqnarray}
&& \dot r =\pm E,\label{mr.1}\\
&& \frac{\ed r}{\ed t}=\pm B(r).\label{mr.2}
\end{eqnarray}
Note that, the sign $+$ ($-$) corresponds to the photons falling
onto the cosmological (event) horizon. Choosing the initial radial distance  $r=r_i$ for $t=\tau=0$, and regarding the expression in Eq.~\eqref{eq:lapse_2}, the above two equations provide the solutions
\begin{eqnarray}
&&\tau(r) = \pm \frac{r-r_i}{E},\label{eq:tau(r)-null}\\
&& t(r) = \pm\frac{1}{\gamma(r_{++}-r_+)}\left[r_+\ln\left|
\frac{r-r_+}{r_i-r_+}
\right|
-r_{++}\ln\left|\frac{r_{++}-r}{r_{++}-r_i}\right|
\right].
\end{eqnarray}
In Fig.~\ref{fig:time(r)-null}, the above solutions have been demonstrated, which indicate that, whereas the comoving observers see themselves passing both of the horizons, to the distant observers, the geodesics never pass the horizons. 

\subsection{Time-like geodesics}\label{subsec:radial-timelike}

\begin{figure}[t]
	\begin{center}
		\includegraphics[width=8cm]{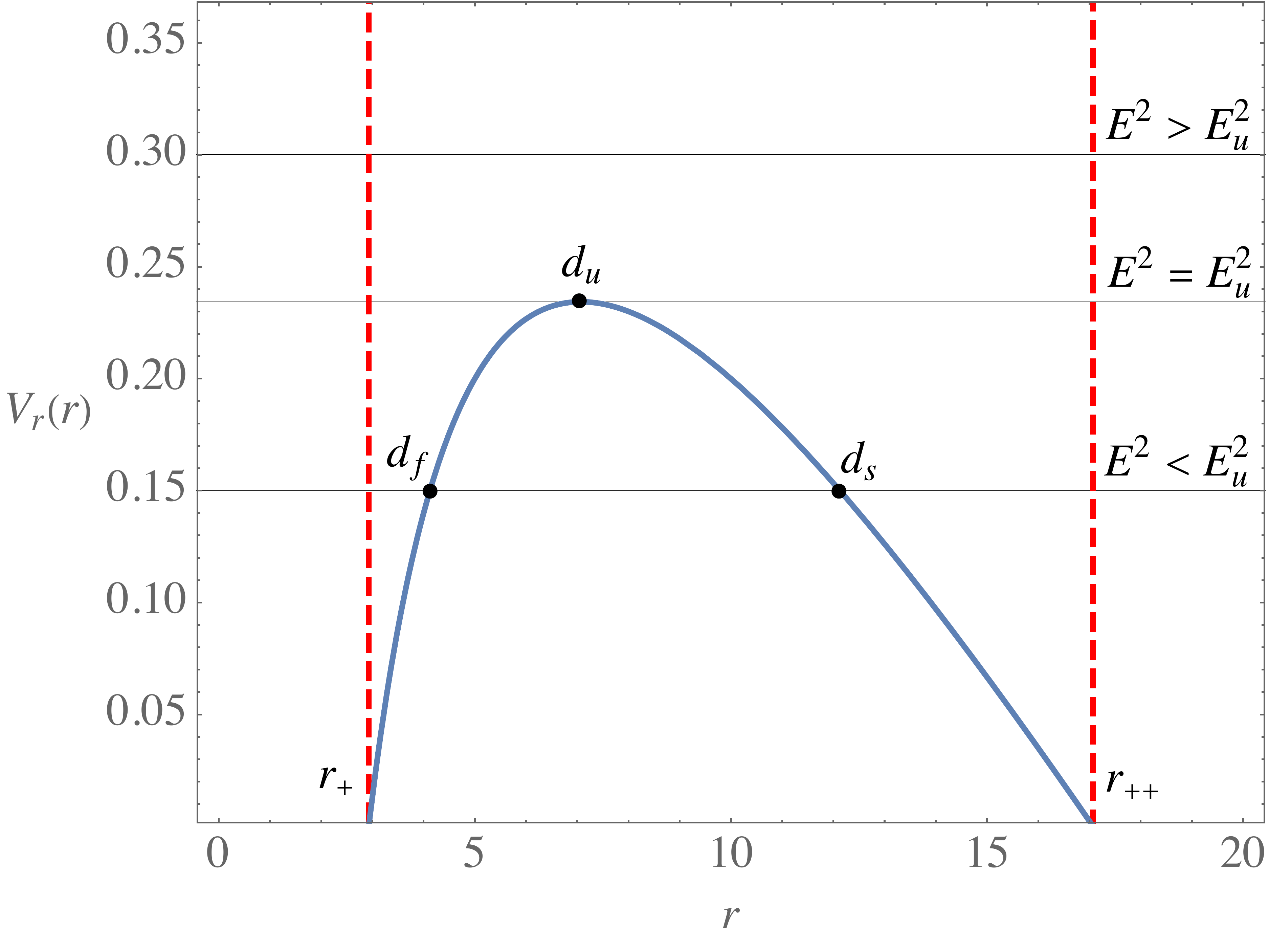}
	\end{center}
	\caption{The effective potential for radially moving massive particles plotted for $\gamma=0.04$. In this particular case, the maximum distance of unstable orbit, is $d_u=7.07$ in accordance with $E^2=E^2_u=0.23$, and the two turning points $d_s=12.3$ and $d_f=4.12$ have been indicated in accordance with the corresponding value $E^2=0.15$. The point $d_s$ is related to the distance, at which, the particles $0<E^2<E^2_u$, experience their frontal scattering. }
	\label{fig:Veff-Radial_timeLike}
\end{figure}

As discussed in the previous subsection, to the distant observers, radially infalling test particles never cross the horizons and in this sense, they appear \textit{frozen} (see Refs.~\cite{Ryder:2009,Zeldovich:2014}). Such particles confront an effective potential $V_r(r)$ that is obtained by letting $\epsilon = 1$ and $L=0$ in Eq.~\eqref{eq:Veff}, whose profile has been shown in Fig. \ref{fig:Veff-Radial_timeLike}.
Accordingly, the motion becomes unstable where $V'_r(r)=0$, solving which, yields 
\begin{equation}\label{rt1} 
d_u=\sqrt{\frac{2M}{\gamma}},
\end{equation} 
as the maximum distance of the unstable motion. Taking into account $E_u^2 \equiv V_{r}(d_u)=1-\alpha -2\sqrt{2M\gamma}$, the possible radial orbits are then categorized as follows, based on the value of $E$ compared with its critical value $E_u$:
\begin{itemize}

	\item \textit{Frontal scatterings of the first and the second kind (FSFK and FSSK)}: For $0<E^2<E^2_u$, the potential allows for the turning point $d_s$ ($d_u<d_s<r_{++}$) corresponding to the scattering distance (FSFK). In the case of $0<E^2<E^2_u$, another turning point $d_f$ ($r_+<d_f<d_{u}$) occurs for those trajectories that fall onto the event horizon and are captured (RSSK).
	
		\item \textit{Critical radial motion}: For $E^2=E^2_u$, the test particles that come from the initial distance $d_i$ ($d_u < d_i < r_{++}$) fall on the unstable  radius $d_u$, and those that come from $d_0$  ($r_+ < d_0 < d_u$), cross the horizon.
		
		\item \textit{Radial capture}: For $E^2>E^2_u$, particles approaching a finite distance $d_j$ ($r_+ < d_j< r_{++}$),  fall directly onto the event horizon. 
		
\end{itemize} 
Now, to formulate the time-like radial orbits, let us recall the radial velocity relations
\begin{eqnarray}
&& \dot{r}^{2}=\frac{\gamma \, \mathfrak{p}(r)}{r},
\label{vel1}\\
&&\left(\frac{\ed r}{\ed t}\right)^{2}=\frac{\gamma^3
	(r-r_{+})^2(r_{++}-r)^2 \mathfrak{p}(r)}{E^2 r^3},\label{vel2}
\end{eqnarray}
which are inferred from Eqs.~\eqref{eq:lapse_2}, \eqref{eq:rdot} and \eqref{rt}, with
\begin{equation}\label{eq:polyRadial-timelike}
\mathfrak{p}(r)\equiv  r^2-\left(\frac{1-\alpha-E^2}{\gamma} \right)  r+\frac{2M}{\gamma}.
\end{equation}
In general, the equation $\mathfrak{p}(r)=0$ is satisfied at any turning point for the radial orbits. We continue by discussing the frontal scatterings.

\subsubsection{FSFK and FSSK}\label{subsubsec:FSFK}

\begin{figure}[t]
	\begin{center}
		\includegraphics[width=7.5cm]{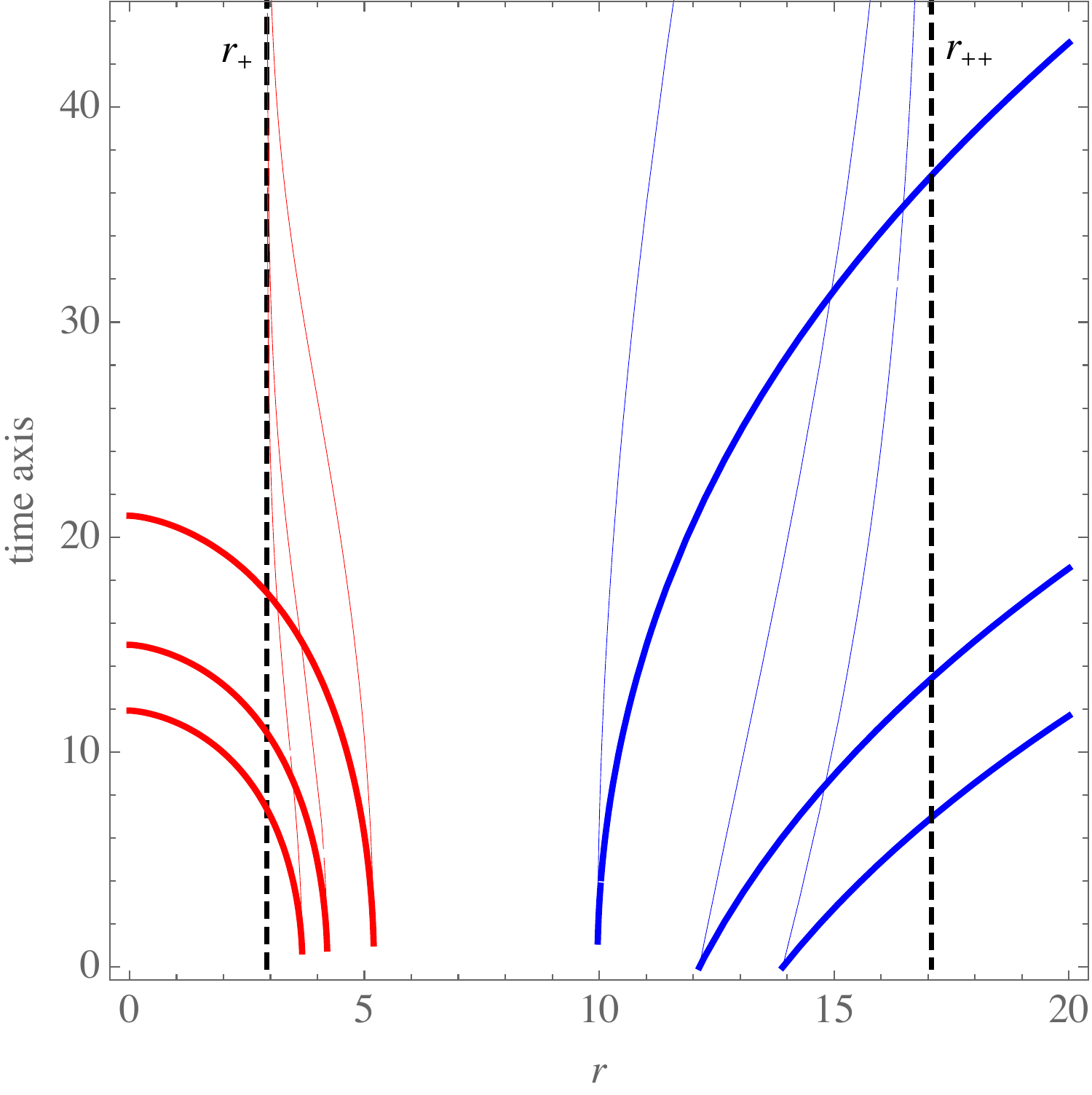}
	\end{center}
	\caption{The plots of FSFK (blue) and FSSK (red), for $\gamma=0.04$ and in accordance with three different energies $E^2=0.1, 0.15$ and $0.21$ (from bottom to top), corresponding to three different initial points for each of the cases (i.e. $d_s$ for FSFK and $d_f$ for FSSK). The thick curves show the radial evolution of the proper time $\tau(r)$, whereas the thin ones demonstrate that of the coordinate time $t(r)$. 
	}
	\label{fig:FS}
\end{figure}

The polynomial \eqref{eq:polyRadial-timelike} vanishes for the two radii
\begin{eqnarray}
&& d_s = \frac{1-\alpha-E^2}{\gamma}\cos^2\left(\frac{1}{2}\arcsin\left(\frac{2\sqrt{2M\gamma}}{1-\alpha-E^2}\right)\right),\label{rs}  \\
&& d_f = \frac{1-\alpha-E^2}{\gamma}\sin^2\left(\frac{1}{2}\arcsin\left(\frac{2\sqrt{2M\gamma}}{1-\alpha-E^2}\right)\right).\label{rs}  
\end{eqnarray}
For the case of FSFK that occurs at $r=r_s$, the differential equation \eqref{vel1} provides the solution
\begin{equation}\label{eq:tau(r)-int-4}
\tau (U)={d_s \over 4\sqrt{\gamma d_f}}\left[  F(U_s)-  F(U)
\right], 
\end{equation}
where 
\begin{equation}\label{eq:F(U)}
F(U)=\frac{1}{\wp'(\Omega_s)}
\left[\ln\left|\frac{\sigma\left(\ss(U)-\Omega_s\right)}
{\sigma\left(\ss(\mathrm{U})+\Omega_s\right)}\right|+2\zeta(\Omega_s)\ss(U)
\right].
\end{equation}
In the above relation, $\wp'(x)\equiv\frac{\ed}{\ed x}\wp(x)$, in which, $\wp(x)\equiv\wp(x;g_2,g_3)$ is the $\wp$-\We{ian} elliptic function with the invariants $g_2$ and $g_3$, whose inverse is notated by $\wp^{-1}(x)\equiv\ss(x)$. Furthermore, $\zeta(x)$ and $\sigma(x)$ are also the Zeta and Sigma \We{ian} functions with the same invariants (for more information, see Ref.~\cite{Handbook:1971}). In Eq.~\eqref{eq:F(U)}
\begin{subequations}\label{radialUUsOmegas}
	\begin{align}
	& U(r)={d_s \over 4r}-{d_s+d_f\over 12 d_f},\label{radialUUsOmegas-a}\\
		& U_s={1 \over 4}-{d_s+d_f\over 12  d_f},\label{radialUUsOmegas-b}\\
	& \Omega_s =\ss\left(-{d_s+d_f\over 12 d_f}\right),\label{radialUUsOmegas-c}\\
	\end{align}
\end{subequations}
and the corresponding \We{} invariants are
\begin{subequations}\label{radialg2g3}
	\begin{align}
	& g_2 = \frac{1}{12d_f^2}\left(d_s^2-d_sd_f+d_f^2\right),\label{radialg2g3-a}\\
	& g_3 =\frac{1}{432d_f^3}(d_s-2d_f)(d_s-d_f)(2d_s-d_f).\label{radialg2g3-b}
	\end{align}
\end{subequations}
The solution in Eq.~\eqref{eq:tau(r)-int-4} corresponds to the radial evolution of the proper time for comoving observers. For the case of distant observers, one can integrate Eq.~\eqref{vel2}, that yields
\begin{equation}\label{teradsct}
t(U)=	\delta_{0}\sum_{k=1}^{2} \delta_k\left[F_k(U_s)- F_k(U)\right] ,
\end{equation}
with the same expressions for $U(r)$ and $U_s$ as in Eqs.~\eqref{radialUUsOmegas-a} and \eqref{radialUUsOmegas-b}. In the above solution, we have defined
\begin{equation}\label{eq:Fj(U)}
F_k(U) = \frac{1}{\wp'(\Omega_k)}\left[\ln\left|\frac{\sigma\left(\ss(U)-\Omega_k\right)}
{\sigma\left(\ss(U)+\Omega_k\right)}\right|+2\zeta(\Omega_k)\ss(U)
\right],
\end{equation}
for which the \We{} invariants are those in Eqs.~\eqref{radialg2g3}, and 
\begin{subequations}\label{eq:d0d1d2O1O2}
	\begin{align}
&	\delta_{0} = {E d_s \over 4\gamma\sqrt{\gamma d_f}(r_{++}-r_+)},\label{eq:d0d1d2O1O2-a}\\
&	\delta_{1} =  1,\label{eq:d0d1d2O1O2-b}\\
&	\delta_{2} =   -1,\label{eq:d0d1d2O1O2-c}\\
&	\Omega_1 =\ss\left({d_s\over 4 r_{++}}-{d_s+d_f\over 12 d_f}\right),\label{eq:d0d1d2O1O2-d}\\
&  \Omega_2  =\ss\left({d_s\over 4 r_{+}}-{d_s+d_f\over 12 d_f}\right).\label{eq:d0d1d2O1O2-e}
	\end{align}
\end{subequations}
The above solutions describe radially moving particles on the FSFK, which are scattered at the distance $d_s$. Accordingly, to obtain the solutions for the FSSK, it is  enough to do the change $d_s \rightarrow d_f$ in the above solutions, and reverse the direction of the radial propagation. The frontal scatterings of time-like geodesics have been shown in Fig.~\ref{fig:FS}, for both of the FSFK and FSSK. 

\subsubsection{Critical radial motion}

\begin{figure}[t]
	\begin{center}
		\includegraphics[width=7.5cm]{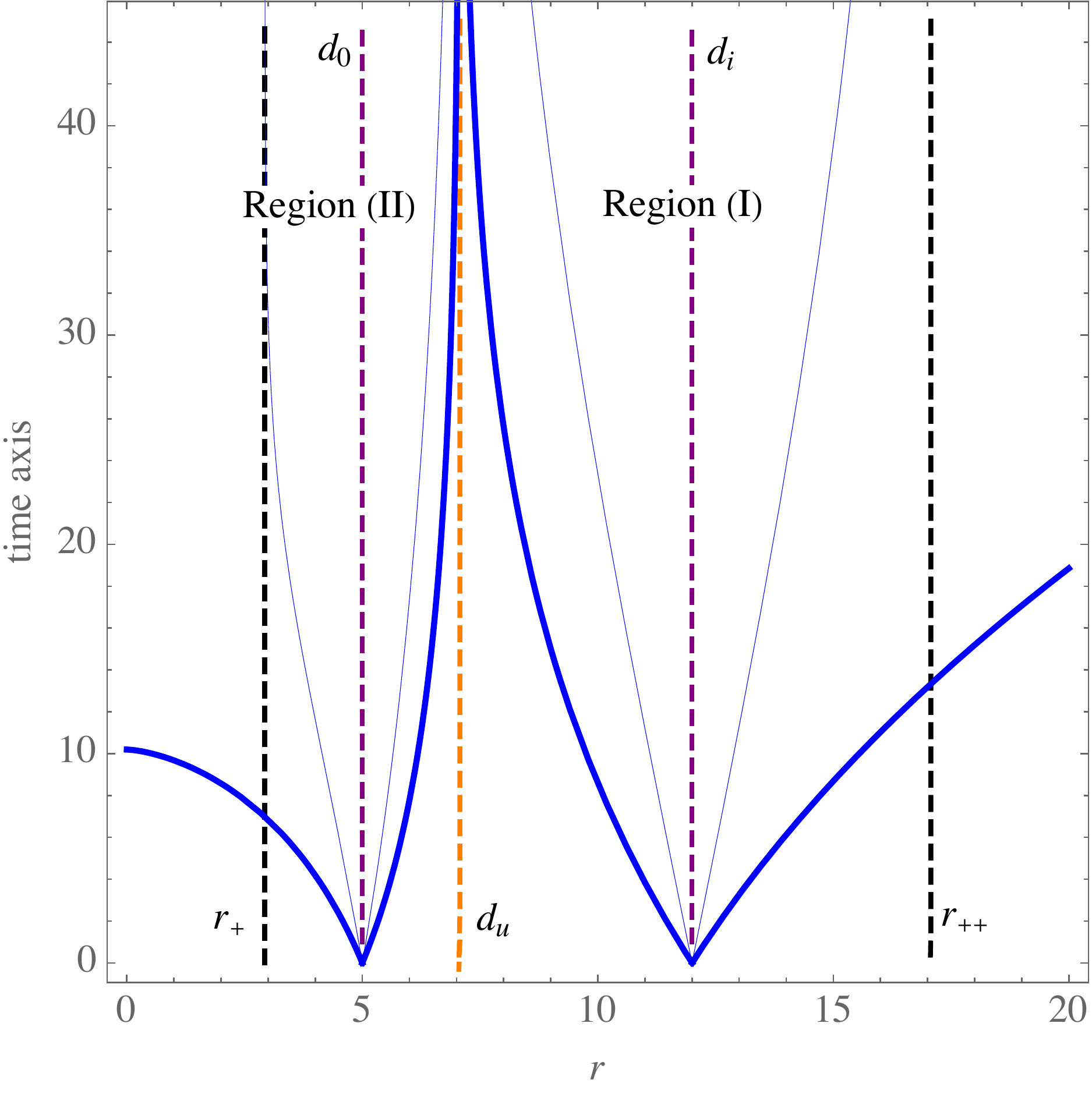}
	\end{center}
	\caption{The critical radial motion in regions (I) and (II), plotted for comoving (thick curves) and distant (thin curves) observers, for the case of $\gamma=0.04$. This value provides $d_u = 7.07$ corresponding to $E_u^2=0.23$. The trajectories have been specified for particles approaching from $d_0=5$ and $d_i=12$.}
	\label{fig:radialCritical-timelike}
\end{figure}

Particles with $E^2 = E^2_u$, can approach from two directions, initiating from either the radial distance $d_i$ ($d_u<d_i<r_{++}$), or from $d_0$ ($r_+<d_0<d_u$), each of which, lead to a different fate. We distinguish these fates, respectively by regions (I) and (II), as indicated in Fig.~\ref{fig:radialCritical-timelike}. Solving the temporal equation \eqref{vel1}, and by taking into account the fact that for critical orbits the characteristic polynomial changes its form to $\mathfrak{p}(r)=(r-d_u)^2$, we then obtain
\begin{subequations}
\begin{align}
& \tau_{\mathrm{I}} (r) = \pm  \left[ \tau_A (r,d_u)-\tau_A (d_i,d_u)\right],  \label{tauc1}\\
& \tau_{\mathrm{II}} (r) = \mp  \left[  \tau_A (r,d_u)-\tau_A (d_0,d_u)\right],\label{tauc2}
\end{align}
\end{subequations}
where we have defined the function
\begin{equation}\label{eq:mathfrakp}
\tau_A(r,y)=2 \sqrt{\frac{r}{\gamma }}-2 \sqrt{\frac{y}{\gamma }} \arctanh \left(\sqrt{\frac{r}{y}}\right).
\end{equation}
For the distant observes and to obtain the evolution of the $t$-coordinate, we integrate Eq.~\eqref{vel2}, which results in 
\begin{subequations}
\begin{align}
t_\mathrm{I} (r) = \pm E_u\sum_{n=1}^{3}\,\varpi_n\,\left[t_n(r)-t_n(d_i)\right], \label{tc1}\\
t_{\mathrm{II}} (r) = \mp E_u \sum_{n=1}^{3} \varpi_n \left[t_n(r)-t_n(d_0)\right],\label{tc2}
\end{align}
\end{subequations}
with
\begin{subequations}
	\begin{align}
&   t_1(r)= \tau_A(r,r_{++}),\\
&	t_2(r)= \tau_A(r,r_{+}),\\
&	t_3(r)= \tau_A(r,d_u),
	\end{align}
\end{subequations}
and
\begin{subequations}
	\begin{align}
&	\varpi_{1} = {-r_{++}\over \gamma \,(r_{++}-r_{+})(r_{++}-d_u)},\\
&	\varpi_{2} = {-r_{+}\over \gamma \,(r_{++}-r_{+})(d_u-r_{+})},\\
&	\varpi_{3} = {d_u\over \gamma \,(r_{++}-d_u)(d_u-r_{+})}.
\end{align}
\end{subequations}
The critical radial motion of the temporal coordinates has been plotted in Fig.~\ref{fig:radialCritical-timelike}, in accordance with the above solutions and for the two different initial distances $d_i$ and $d_0$, that generate the discussed regions.

\section{Angular motion}\label{sec:angular}

Particles with $L\neq0$ can obtain more general types of geodesic motion. In this section, we analyze the angular motion of mass-less and massive particles, by solving the angular equation of motion \eqref{rphi}. As before, we apply the methods of integrating the elliptic integrals that appear in the course of this study. In a spacial occasion, a hyper-elliptic integral is generated for the case of planetary orbits of massive particles. This integral will be dealt with by means of a particular form of the Lauricella hypergeometric function.

\subsection{Null geodesics}\label{subsec:NullAngular}

\begin{figure}[t]
	\begin{center}
		\includegraphics[width=8cm]{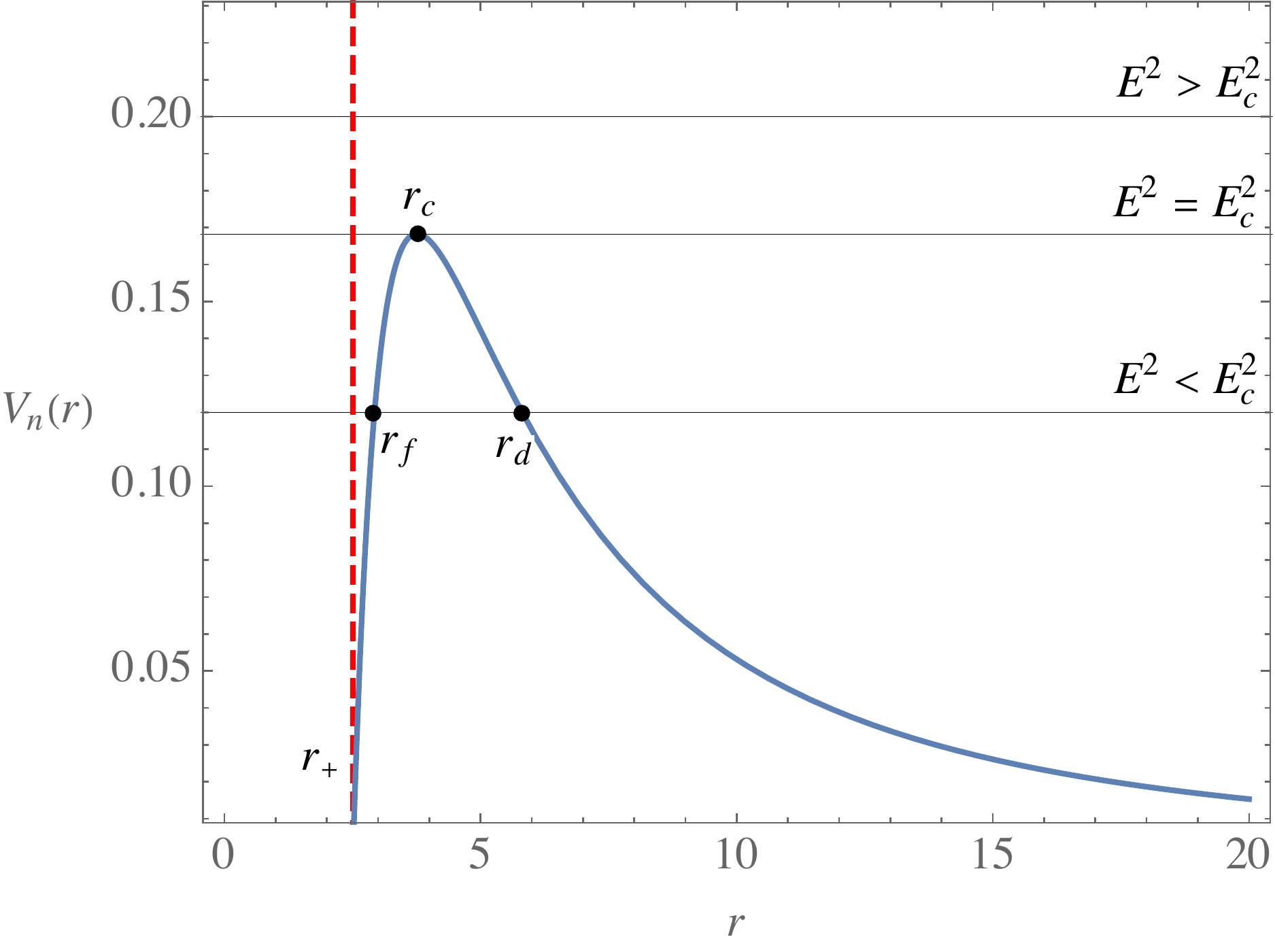}
	\end{center}
	\caption{The effective potential for angular null geodesics plotted for $\gamma=0.001$ and $L=3$. Altering the values of $L$ only affects the scale of the potential and leaves its shape unchanged. For this particular potential, the turning points are located at $r_d=5.81$ and $r_f=2.93$ corresponding to $E^2=0.12$, and $r_c=3.76$ corresponding to $E_c^2=0.168$.}
	\label{fig:Vn}
\end{figure}

For the case of $\epsilon=0$ in Eq.~\eqref{eq:Veff}, the effective potential for the angular null geodesics becomes
\begin{equation}
V_n(r) = \frac{\gamma L^2}{r^3}\left(r-r_+\right)\left(r_{++}-r\right),
    \label{eq:Vn}
\end{equation}
which has been plotted in Fig.~\ref{fig:Vn}. It can be checked that raising and lowering $L$, changes only the scale of the axes, not the shape of the effective potential. So, the one given in Fig.~\ref{fig:Vn}, can categorize all the possible null orbits. As before, the turning points are where $E^2 = V_n(r_t)$ is satisfied, which for the case of angular geodesics, correspond to the vanishing of the angular equation of motion \eqref{rphi}.
In Fig.~\ref{fig:Vn}, these turning points have been indicated by $r_d$ and $r_f$, indicating respectively, the radial distances for deflecting trajectories or orbits of the first kind (OFK) and terminating bound orbits or obits of the second kind (OSK). The other turning point $r_c$, at which the condition $V'(r_c)=0$ is satisfied, provides two kinds of unstable orbits, termed as the critical orbits of the first and second kind (COFK and COSK).

\subsubsection{Period of the unstable circular orbits}\label{subsubsec:periodofrc}

The condition $V'_n(r_c) = 0$ results in the radial distance
 \begin{equation}\label{ptu}
  r_{c}=\frac{2(1-\alpha)}{\gamma}\sin^2\left(\frac{1}{2}\arcsin\left(\frac{\sqrt{6M\gamma}}{1-\alpha}\right)\right).
  \end{equation}
 To compute the proper and the coordinate periods of orbits at $r_c$, we consider the expressions in Eqs.~\eqref{eq:cteMotion}, from which, we infer
 \begin{eqnarray}
 && \Delta \tau_c=\frac{r_c^2}{L}\Delta\phi_c,\label{eq:tauc}\\
 && \Delta t_c=\frac{E_c}{L}\frac{r_c^2}{B(r_c)}\Delta\phi_c.\label{eq:tc}
 \end{eqnarray}
 This way, for one complete revolution, one obtains the periods
  \begin{equation}\label{p1}
 \left. T_{\tau}\equiv \Delta\tau_c \right|_{\Delta\phi_c=\frac{\pi}{2}}= \frac{2\pi r_c^2}{L},
  \end{equation}
 for comoving observers and 
  \begin{equation}\label{p2}
\left. T_{t}\equiv \Delta t_c \right|_{\Delta\phi_c=\frac{\pi}{2}}= \frac{2\pi r_c}{\sqrt{B(r_c)}},
  \end{equation}
for distant observers.

\subsubsection{Deflecting trajectories}\label{subsubsec:deflection-light}

The turning points $r_d$ and $r_f$ can be calculated by solving the equation
\begin{equation}
\left(\frac{\ed r}{\ed\phi}\right)^2 = \frac{\mathcal{P}_4(r)}{L^2} =0,
    \label{eq:character-null}
\end{equation}
where $\mathcal{P}_4(r) = r\left[ E^2r^3+L^2\gamma r^2-L^2(1-\alpha)r+2ML^2\right]$. Beside the trivial solution $r=0$, the equation $\mathcal{P}_4(r)=0$ has the three real roots
\begin{eqnarray}
    && r_d = \sqrt{\frac{\xi_2}{3}}\cosh\left(
    \frac{1}{3}\arccosh\left(3\xi_3\sqrt{\frac{3}{\xi_2^3}}\right)
    \right)-\frac{\gamma b^2}{3},\label{eq:rd-null}\\
    && r_n = \sqrt{\frac{\xi_2}{3}}\cosh\left(
    \frac{1}{3}\arccosh\left(3\xi_3\sqrt{\frac{3}{\xi_2^3}}\right)+\frac{2\pi\im}{3}
    \right)-\frac{\gamma b^2}{3},\label{eq:rn-null}\\
    && r_f = \sqrt{\frac{\xi_2}{3}}\cosh\left(
    \frac{1}{3}\arccosh\left(3\xi_3\sqrt{\frac{3}{\xi_2^3}}\right)+\frac{4\pi\im}{3}
    \right)-\frac{\gamma b^2}{3},\label{eq:rf-null}
\end{eqnarray}
in which, as expected, the positive-valued ones are only $r_d$ and $r_f$, and $b\equiv\frac{L}{E}$ is the impact parameter. Here, we have defined
\begin{subequations}
\begin{align}
    & \xi_2= 4\left[\frac{\gamma b^2}{3}+b^2(1-\alpha)\right],\label{eq:xi2}\\
    & \xi_3=-4\left[\frac{\gamma b^4}{3}(1-\alpha)+\frac{2\gamma^3 b^6}{27}+2b^2M\right].\label{eq:xi3}
\end{align}
\end{subequations}
This way, one can recast $\mathcal{P}_4(r)=\frac{E^2}{4}r(r-r_d)(r-r_f)(r-r_n)$, which eases the intergation of the equation of motion. To obtain the  analytical solution for the deflecting trajectories, we first consider the OFK at $r_d$. The direct intergation of the relation $L\left(\frac{\ed r}{\ed\phi}\right) = \sqrt{\mathcal{P}_4(r)}$, results in 
\begin{equation}
r(\phi) = \frac{r_f r_n}{r_d \left[\frac{\ell_0}{3 r_d^2}-4\wp\left(\omega_d-\kappa_d\phi\right)\right]},
    \label{eq:rd(phi)-null}
\end{equation}
with the \We{} invariants
\begin{subequations}\label{eq:tg}
\begin{align}
    & \tilde{g}_2 = \frac{\ell_0^2-3r_fr_n\ell_1}{12r_d^4},\label{eq:tg2}\\
    & \tilde{g}_3 = -\frac{27 (r_d r_f r_n)^2+2\ell_0^3-9r_fr_n\ell_0\ell_1}{432 r_d^6},\label{eq:tg3}
\end{align}
\end{subequations}
where $\omega_d=\ss(U_d)$, in which
\begin{equation}
U_d=\frac{\ell_0-3r_f r_n}{12 r_d^2},
    \label{eq:Ud-null}
\end{equation}
and we have defined the constants
\begin{subequations}\label{eq:coefficients-null-def}
\begin{align}
    & \kappa_d=\frac{r_f r_n}{8b r_d},\label{eq:kappad}\\
    & \ell_0=r_d r_f + r_d r_n+r_f r_n,\label{eq:ell0}\\
    & \ell_1=r_d^2+r_d r_f+r_d r_n.\label{eq:ell1}
\end{align}
\end{subequations}
\begin{figure}[t]
	\begin{center}
		\includegraphics[width=7cm]{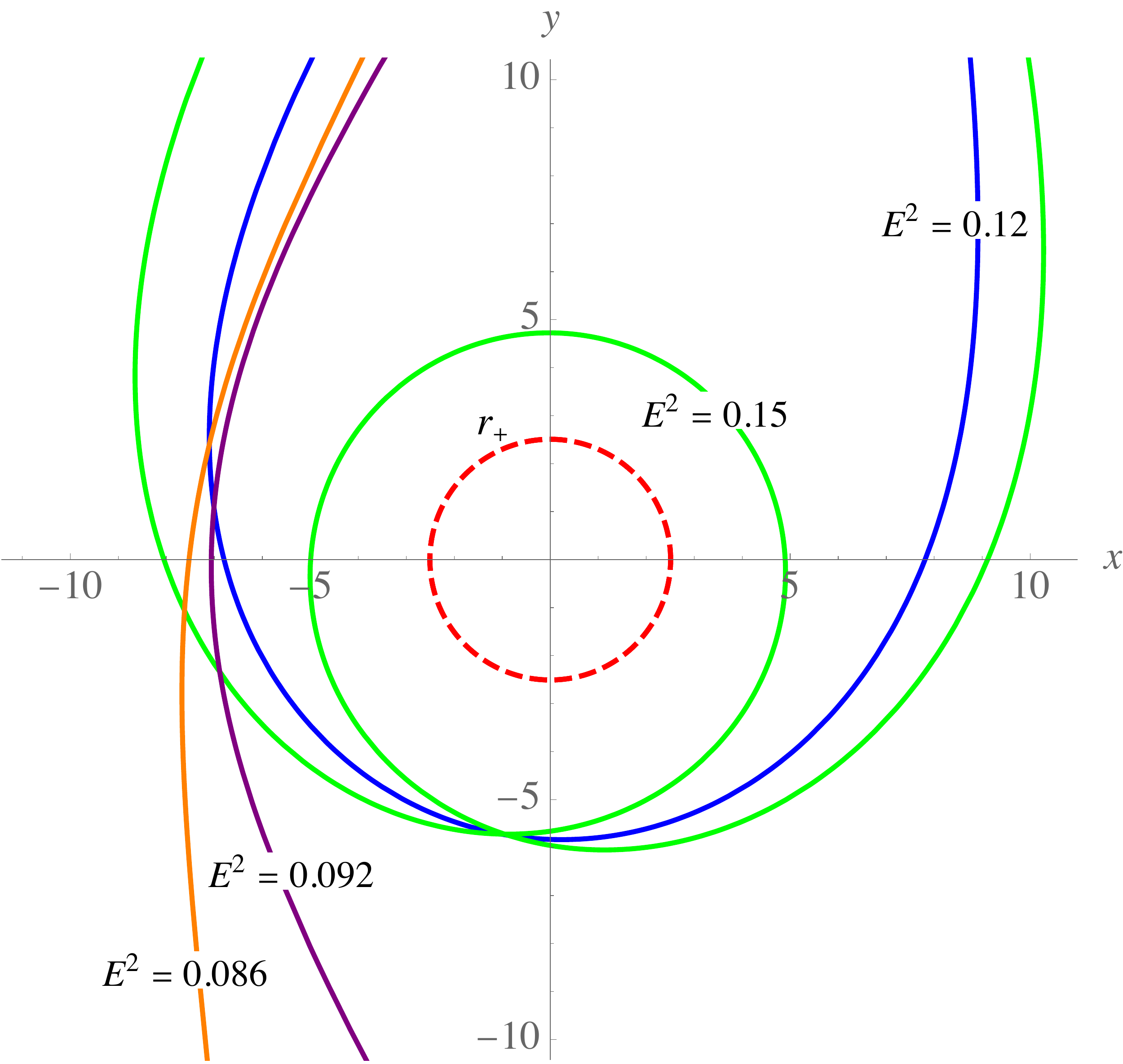}~(a)
		\includegraphics[width=6cm]{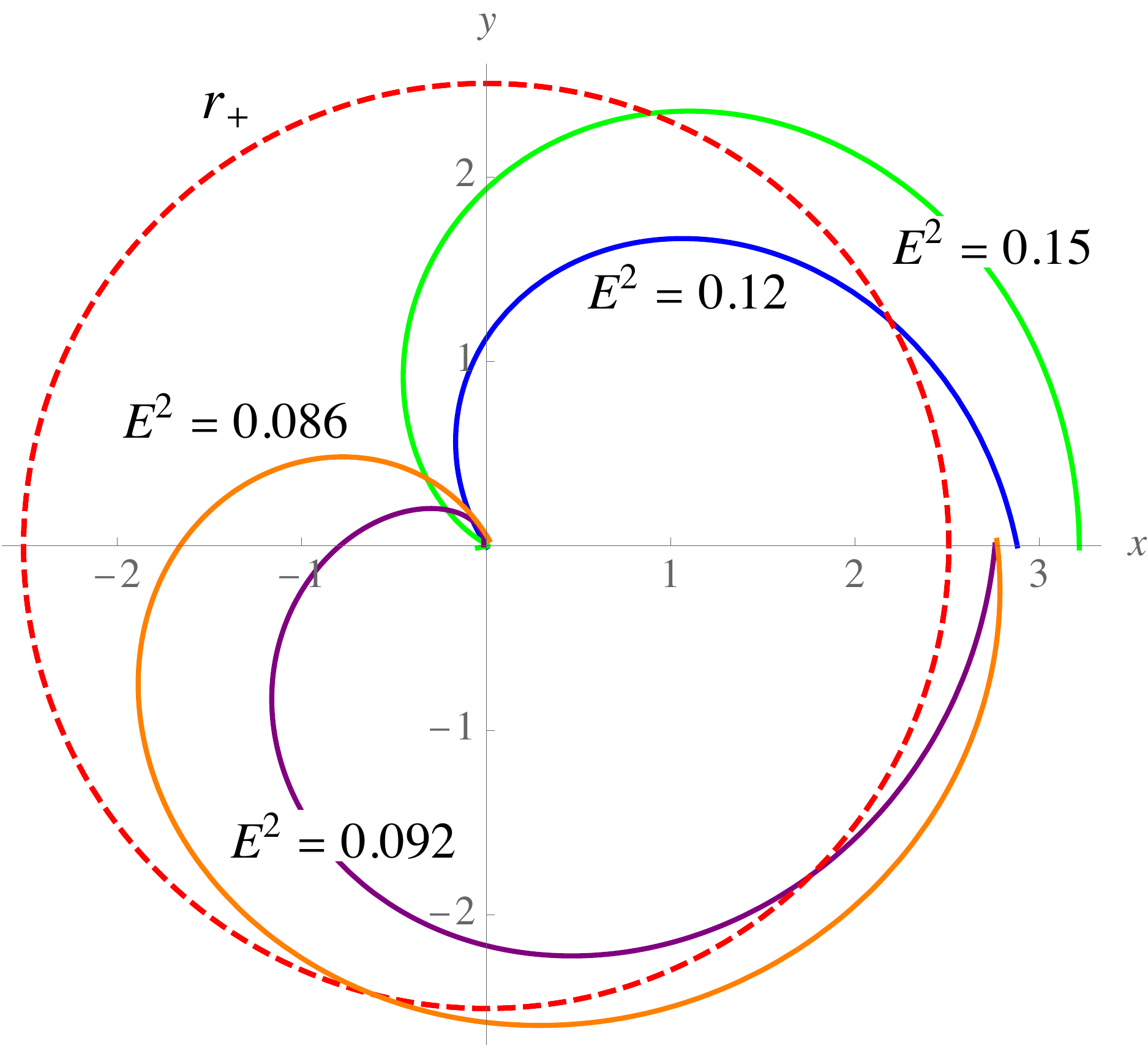}~(b)
	\end{center}
	\caption{The deflecting trajectories for null geodesics corresponding to (a) OFK, and (b) OSK, plotted for different values of $E^2$ and in accordance with the parameters given in Fig.~\ref{fig:Vn}.}
	\label{fig:OFSK-null}
\end{figure}
Pursuing the same procedure for photons approaching from $r_f$, one obtains the analytical solution
\begin{equation}
r(\phi) = \frac{r_d r_n}{r_f \left[\frac{\ell_0}{3 r_f^2}-4\wp\left(\omega_f+\kappa_f\phi\right)\right]},
    \label{eq:rf(phi)-null}
\end{equation}
in which, the \We{} invariants and the constants $\omega_f$ and $\kappa_f$ have the same forms as in Eqs.~\eqref{eq:tg}--\eqref{eq:coefficients-null-def}, assuming the exchange $r_d\leftrightarrow r_f$. In Fig.~\ref{fig:OFSK-null}, the OFK and OSK have been plotted for several energies.

Note that, the OFK can be used to obtain the lens equation. Accordingly, the angle of deflection due to the gravitational lensing is calculated as \cite{Misner:1973}
\begin{equation}
\hat\vartheta = 2\left(\phi_\infty-\phi_d\right)-\pi,
    \label{eq:vartheta-null}
\end{equation}
in which
$\phi_\infty \equiv\phi(\infty)$ and $\phi_d=\phi(r_d)$, where the radial behavior of the azimuth angle in the equatorial plane is given as
\begin{equation}
    \phi(r) = L\int\frac{\ed r}{\sqrt{\mathcal{P}_4(r)}},
\end{equation}
according to Eq.~\eqref{eq:character-null}. This way, the lens equation is obtained as
\begin{equation}
\hat{\vartheta} = \frac{2}{\kappa_d}\left[
\ss\left(\frac{1}{4r_d^2}\left[
\frac{\ell_0}{3}-r_f r_n
\right]\right)
-\ss\left(\frac{\ell_0}{12 r_d^2}\right)
\right]-\pi,
    \label{eq:lensEq}
\end{equation}
by taking into account the inversion of Eq. \eqref{eq:rd(phi)-null} for an unbound (flyby) orbit.

\subsubsection{Critical trajectories}\label{subsubsec:critical-null}

\begin{figure}[t]
	\begin{center}
		\includegraphics[width=8.5cm]{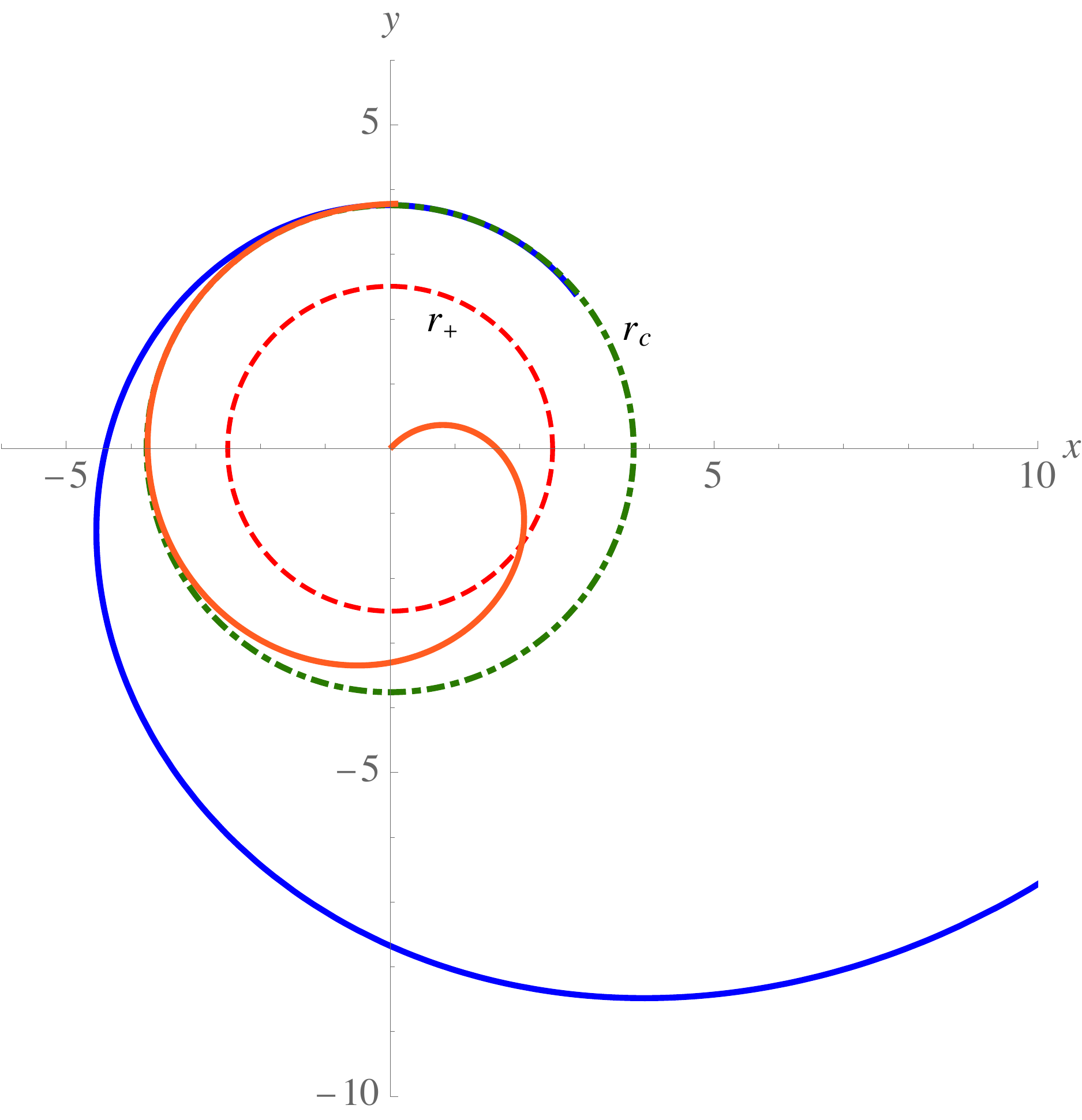}
	\end{center}
	\caption{The COFK (blue) and the COSK (orange), plotted in accordance with the information given in Fig.~\ref{fig:Vn}.
	}
	\label{fig:critical_null}
\end{figure}

In the case of $E=E_c$, the characteristic polynomial gains the form $\mathcal{P}_4(r) = r(r-r_c)^2(r-r_n)$, according to which, one can do the intergation of the equation of motion, based on the approaching points $r_c<r_{i_1}<r_{++}$ and $r_+<r_{i_2}<r_c$. 
This way, and taking into account $b=b_c (\equiv\frac{L}{E_c})$, one obtains the two solutions
\begin{equation}
    r_\mathrm{I}(\phi) = r_c-\frac{r_n}{r_c\left[(r_c-r_n)\tanh\left(\varphi_{i_1}-\kappa_c\phi\right)-r_c\right]},
    \label{eq:COFK}
\end{equation}
that corresponds to the COFK for particles approaching from $r_{i_1}$, and 
\begin{equation}
    r_\mathrm{II}(\phi) = r_c+\frac{r_n}{r_c\left[(r_c-r_n)\tanh\left(\varphi_{i_2}-\kappa_c\phi\right)-r_c\right]},
    \label{eq:COSK}
\end{equation}
corresponding to the COSK for particles that approach from $r_{i_2}$. In these expressions
\begin{subequations}
\begin{align}
    & \kappa_c=\frac{r_c}{2}\sqrt{1-\frac{r_n}{r_c}},\label{eq:kappac}\\
    & \varphi_{i_{1,2}} = \arctanh\left(
    \sqrt{\frac{r_c^2 r_{i_{1,2}}-r_n}{r_c (r_c-r_n) r_{i_{1,2}}}}
    \right).\label{eq:varphii12} 
\end{align}
\end{subequations}
In Fig.~\ref{fig:critical_null}  These orbits have been plotted in accordance with the approaching points, which lead to different fates for the trajectories.

\subsection{Time-like geodesics}\label{subsec:angular-timelike}

For the case of $\epsilon=1$, the effective potential \eqref{eq:Veff} gains the form
\begin{equation}
V_t(r) = \frac{\gamma}{r}\left(r-r_+\right)\left(r_{++}-r\right)\left(1+\frac{L^2}{r^2}\right),
    \label{eq:Vt}
\end{equation}
for the angular time-like geodesics. It is however important to note that, unlike the null case, the values of the angular momentum $L$ are crucial in the characterization of the possible orbits for time-like geodesics. In other words, different choices of $L$ can lead to different available orbits, in accordance with the changes in the shape of the effective potential. Therefore, one needs to find the corresponding limiting values of $L$, that characterize $V_t(r)$. 

To elaborate this, we first consider the limit where the points of inflection and extremums coincide. These points are where the equations $V'(r)=0$ and $V''(r)=0$ are satisfied, simultaneously (marginally stable orbits). The corresponding angular momentums are then ramified to $L_\IS$, for which $V_t(r)$ presents a minimum at $r_\IS$ where the innermost stable circular orbit (ISCO) occurs, and $L_\OS$, for which $V_t(r)$ has a minimum at $r_\OS$ where the outermost stable circular orbit (OSCO) happens. Furthermore, for the critical value $L=L_C$, the effective potential represents two maximum of equal energy levels, occurring at the radial distances $r_{C_1}$ and $r_{C_2}$. Based on the above notions, we can categorize the angular momentums as follows:

\begin{itemize}

\item
For $0<L<L_\IS$ an unstable orbit is available without the presence of stable circular orbits.

\item
For $L=L_\IS$ an unstable orbit and ISCO are available.

\item
For $L_\IS<L<L_C$ a stable circular orbit and two unstable orbits are available. In this case, the first maximum is smaller than the second one, so the energy of the unstable orbit at the larger radius, is greater than that at the smaller radius.

\item
For $L=L_{C}$ there are one stable circular orbit, and two unstable orbits of equal energies.

\item
For $L_{C}<L<L_\OS$ there are one stable circular orbit and two unstable orbits. The energy of the unstable orbit at the smaller radius, is greater than that at the larger radius.

\item
For $L=L_\OS$ an unstable orbit and OSCO are available.

\item
For $L> L_\OS$ only an unstable orbit is available.

\end{itemize}

\begin{figure}[t]
\begin{center}
\includegraphics[width=13cm]{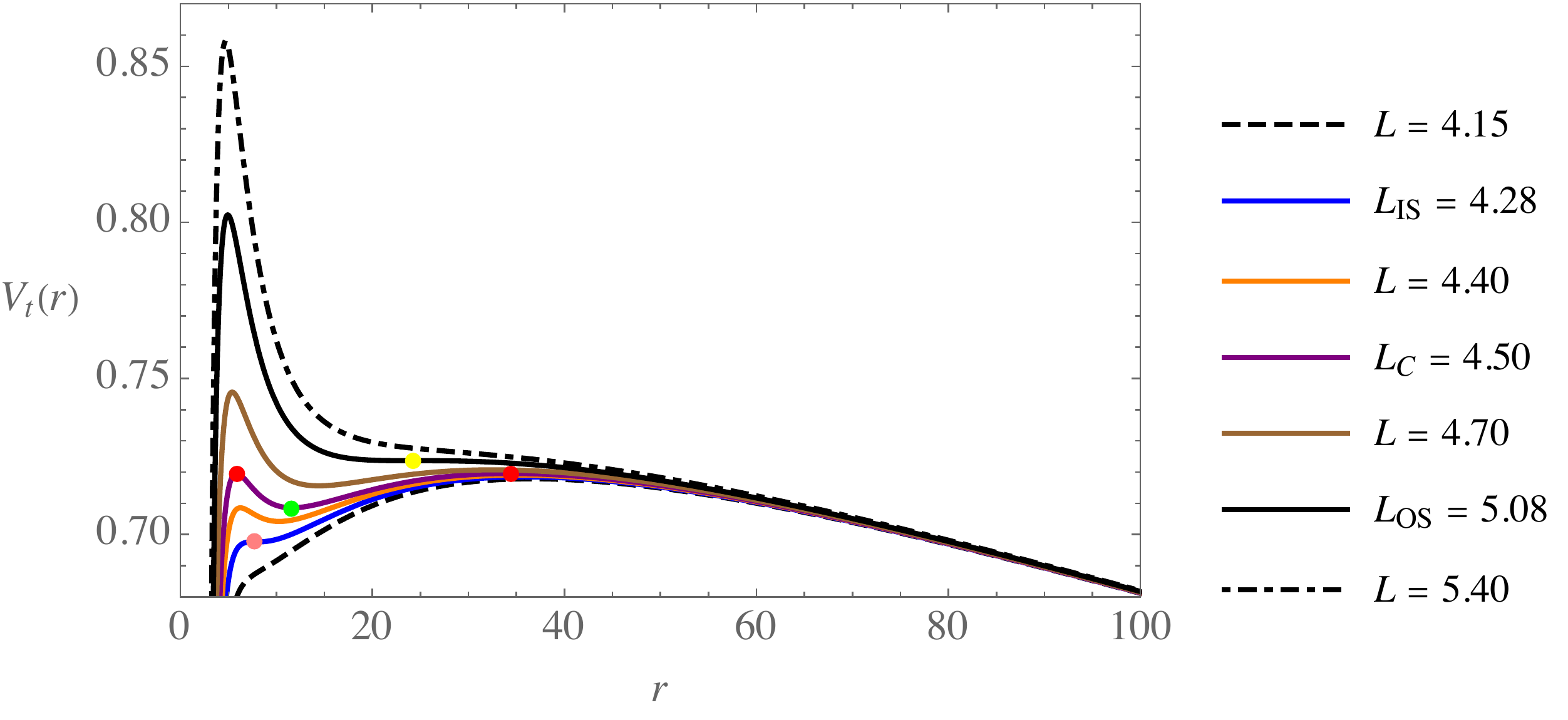}
\end{center}
\caption{The different curves of $V_t(r)$ plotted for $\gamma=0.001$, and ramified in terms of the limiting values of the angular momentum. The red dots indicate the radii of unstable orbits, $r_{C_1}$ and $r_{C_2}$, which are with equal energies and correspond to the $L_C$ curve.} 
\label{fig:Vt}
\end{figure}

In Fig.~\ref{fig:Vt}, several branches of $V_t(r)$ have been plotted, by taking into account the above limits and values.
In what follows, we confine ourselves to the case of $L=L_C$ and the possible time-like orbits are discussed.


\subsubsection{Planetary orbits}\label{subsubsec:planetary}

\begin{figure}[t]
	\begin{center}
		\includegraphics[width=9cm]{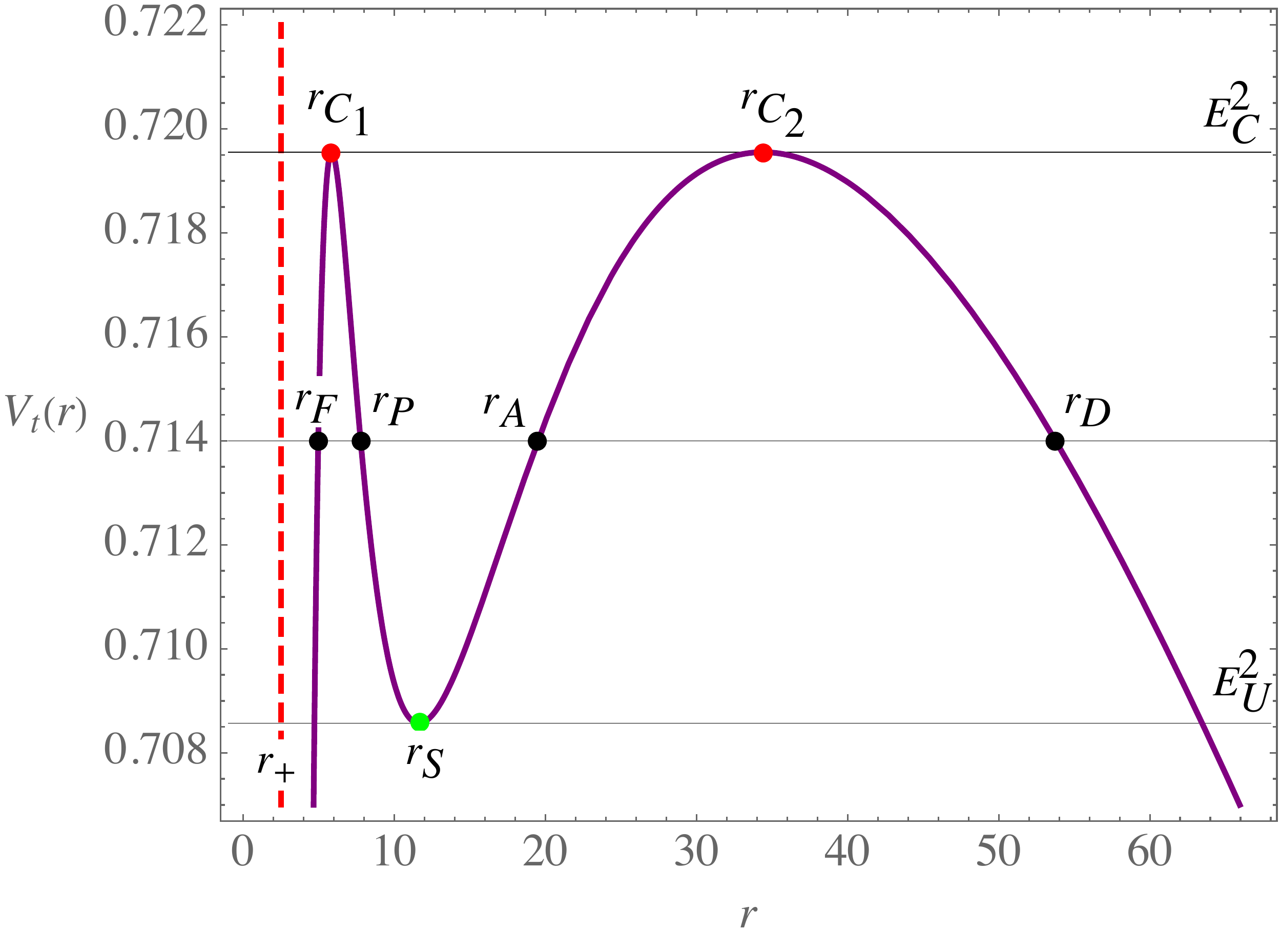}
	\end{center}
	\caption{The behavior of $V_t(r)$ for $L=L_C$ (in accordance with Fig.~\ref{fig:Vt}).  Here, $r_{C_1}=5.85$ and $r_{C_2}=34.37$, corresponding to $E_C^2=0.719$, $r_S=11.63$ corresponding to $E_U^2=0.708$, and $r_D=53.76$, $r_A=19.51$, $r_P=7.77$ and $r_F=4.96$, corresponding to $E^2=0.714$.}
	\label{fig:Vt-planetary}
\end{figure}

Planetary orbits are oscillations between two turning points and are, therefore, bounded between two radii. To discuss the planetary orbits, we select the curve corresponding to $L_C$ from the effective potential, as demonstrated in Fig.~\ref{fig:Vt-planetary}. In this diagram, the stable circular orbits and the critical orbits can occur, respectively, at $r=r_S$ and $r_{C_{1,2}}$. The planetary orbits, however, are only available for the energy level $E_U^2<E^2<E_C^2$, where the 
equation
\begin{equation}
\left(\frac{\ed r}{\ed\phi}\right)^2 = \frac{P_5(r)}{L^2} =0,
    \label{eq:character-timelike}
\end{equation}
in addition to the trivial solution $r=0$, possesses the four real solutions $r_D$ (OFK), $r_A$ (apoapsis), $r_P$ (periapsis), and $r_F$ (OSK). In Eq.~\eqref{eq:character-timelike}, the characteristic polynomial is inferred from Eq.~\eqref{rphi} as $P_5(r) =r\big[\gamma r^4-(1-\alpha-E^2) r^3+(2M+L^2\gamma)r^2-L^2(1-\alpha)r +2ML^2\big]$. The equation $P_5(r)=0$ for this energy level results in the solutions (see appendix \ref{app:A-nube})
\begin{eqnarray}
&& r_D = M\left[ \tilde{\rho}+\sqrt{\tilde{ \rho}^2-\tilde{\beta}}-\frac{\tilde{a}}{4}\right],\label{eq:rD}\\
&& r_A = M\left[ \tilde{\rho}-\sqrt{\tilde{ \rho}^2-\tilde{\beta}}-\frac{\tilde{a}}{4}\right],\label{eq:rA}\\
&& r_P = M\left[ -\tilde{\rho}+\sqrt{\tilde{ \rho}^2-\tilde{\lambda}}-\frac{\tilde{a}}{4}\right],\label{eq:rP}\\
&& r_F = M\left[- \tilde{\rho}-\sqrt{\tilde{ \rho}^2-\tilde{\lambda}}-\frac{\tilde{a}}{4}\right],\label{eq:rF}
\end{eqnarray}
where
\begin{subequations}
\begin{align}
& \tilde{\rho} = \sqrt{\tilde{U}-{\tilde{A}\over 6}},\\
&\tilde{\beta} = 2\tilde{\rho}^2 +{\tilde{A}\over 2}+{\tilde{B}\over 4\tilde{\rho}},\\
& \tilde{\lambda} = 2\tilde{\rho}^2 +{\tilde{A}\over 2}-{\tilde{B}\over 4\tilde{\rho}},
\end{align}
\end{subequations}
with
\begin{subequations}
\begin{align}
& \tilde{A} = \tilde{b}-\frac{3\tilde{a}^2}{8},\\
& \tilde{B} = \tilde{c}+\frac{\tilde{a}^3}{8}-\frac{\tilde{a}\tilde{b}}{2},\\
& \tilde{C} = \tilde{d}+\frac{\tilde{a}^2\tilde{b}}{16}-\frac{3\tilde{a}^4}{256}-\frac{\tilde{a}\tilde{c}}{4},
\end{align}
\end{subequations}
given that
\begin{subequations}
\begin{align}
& \tilde{a} = -\frac{1-\alpha-E^2}{M\gamma  },\\
& \tilde{b} = \frac{2M +L ^2\gamma}{M^2\gamma  },\\
& \tilde{c} = -\frac{L^2(1-\alpha)}{M^3\gamma  },\\
& \tilde{d} = \frac{2L ^2}{M^3\gamma  },
\end{align}
\end{subequations}
and we have defined the function
\begin{equation}
\tilde{U} = \sqrt{{\tilde{\eta}_2\over 3}} \cosh\left(\frac{1}{3}\arccosh\left(3\tilde{\eta}_3\sqrt{{3\over \tilde{\eta}_2^3}}  \right) \right),
    \label{eq:tildeU}
\end{equation}
that includes the constant coefficients
\begin{subequations}
\begin{align}
 &   \tilde{\eta}_2 = {\tilde{A}^2\over 12}+\tilde{C},\\
 &   \tilde{\eta}_3 = {\tilde{A}^3\over 216}-{\tilde{A} \tilde{C}\over 6}+\frac{\tilde{B}^2}{16}.
\end{align}
\end{subequations}
This way, the characteristic polynomial can be recast as $P_5(r) = r(r_D-r)(r_A-r)(r-r_P)(r-r_F)$ for the case that the planetary orbits are present. 

\begin{figure}[t]
	\begin{center}
		\includegraphics[width=4cm]{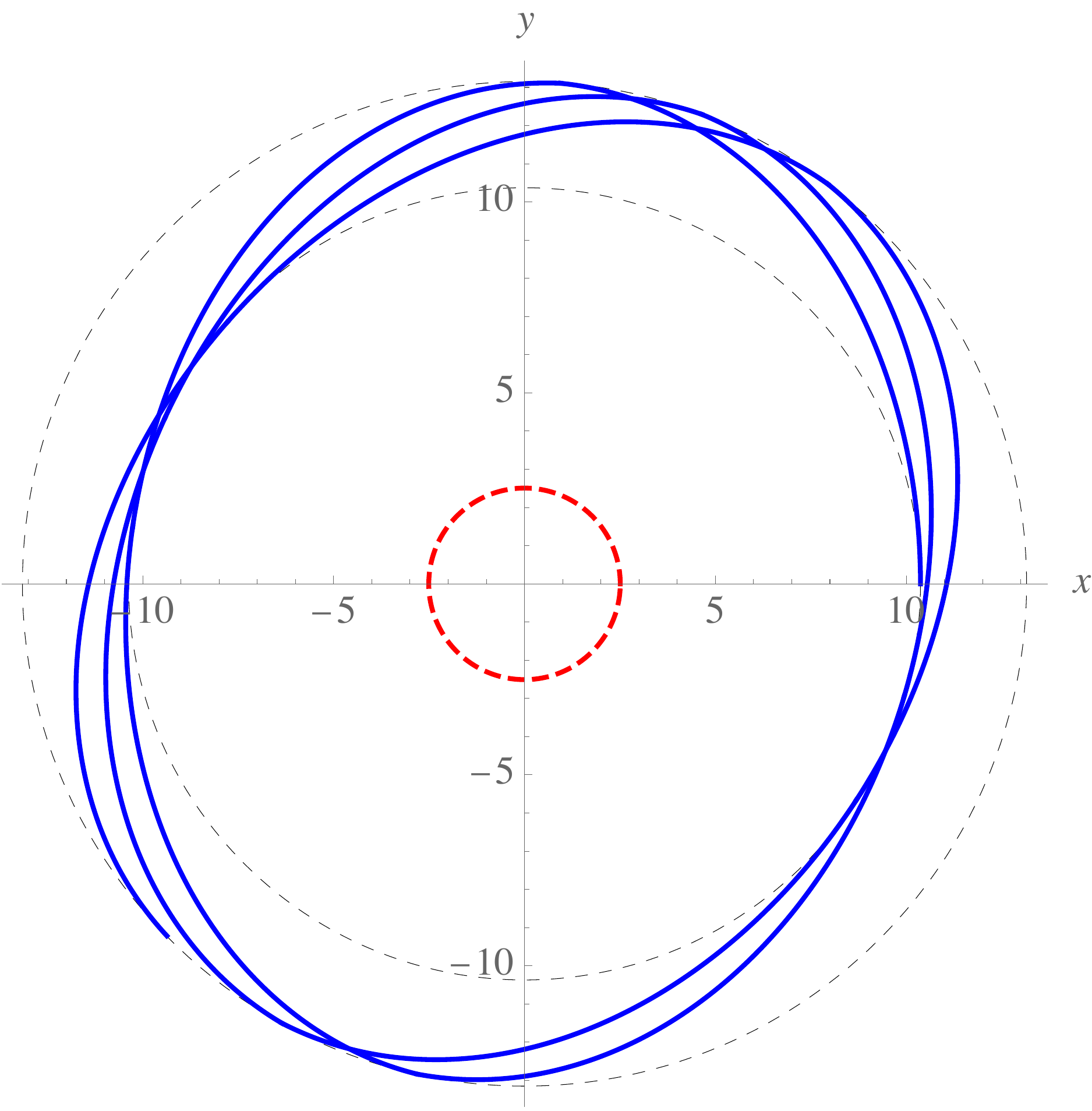}~(a)
		\includegraphics[width=4cm]{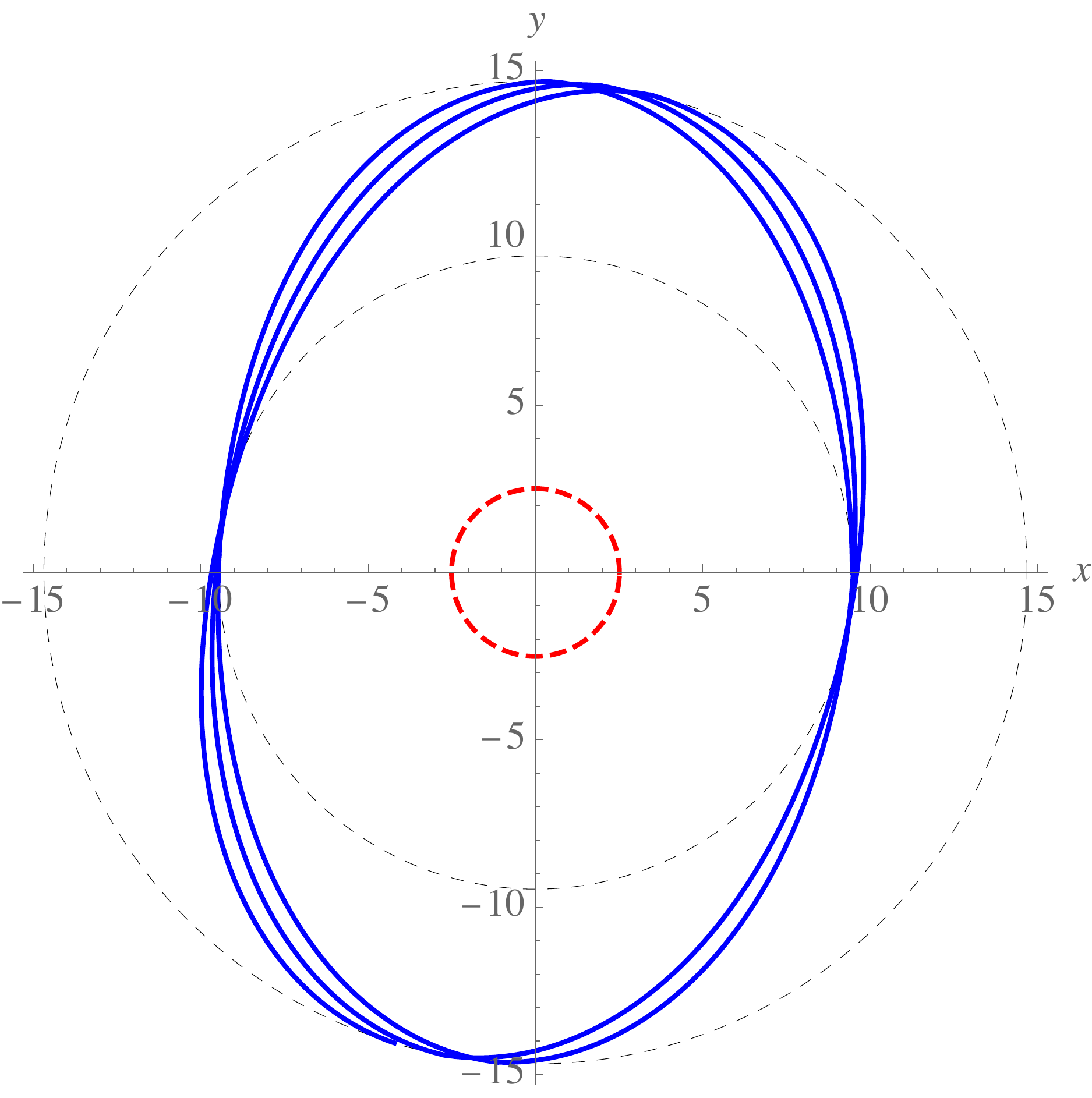}~(b)
		\includegraphics[width=4cm]{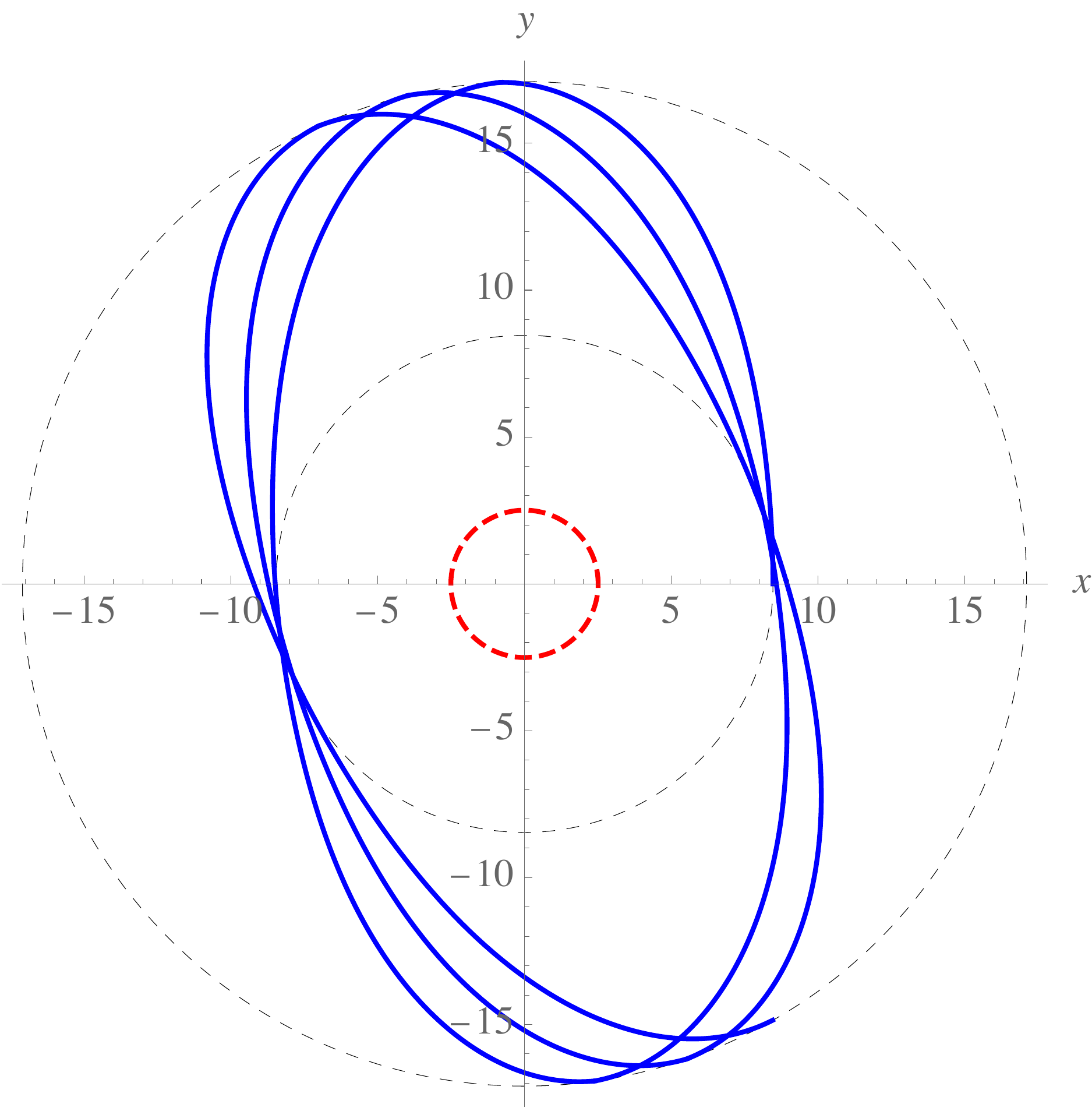}~(c)
		\includegraphics[width=4cm]{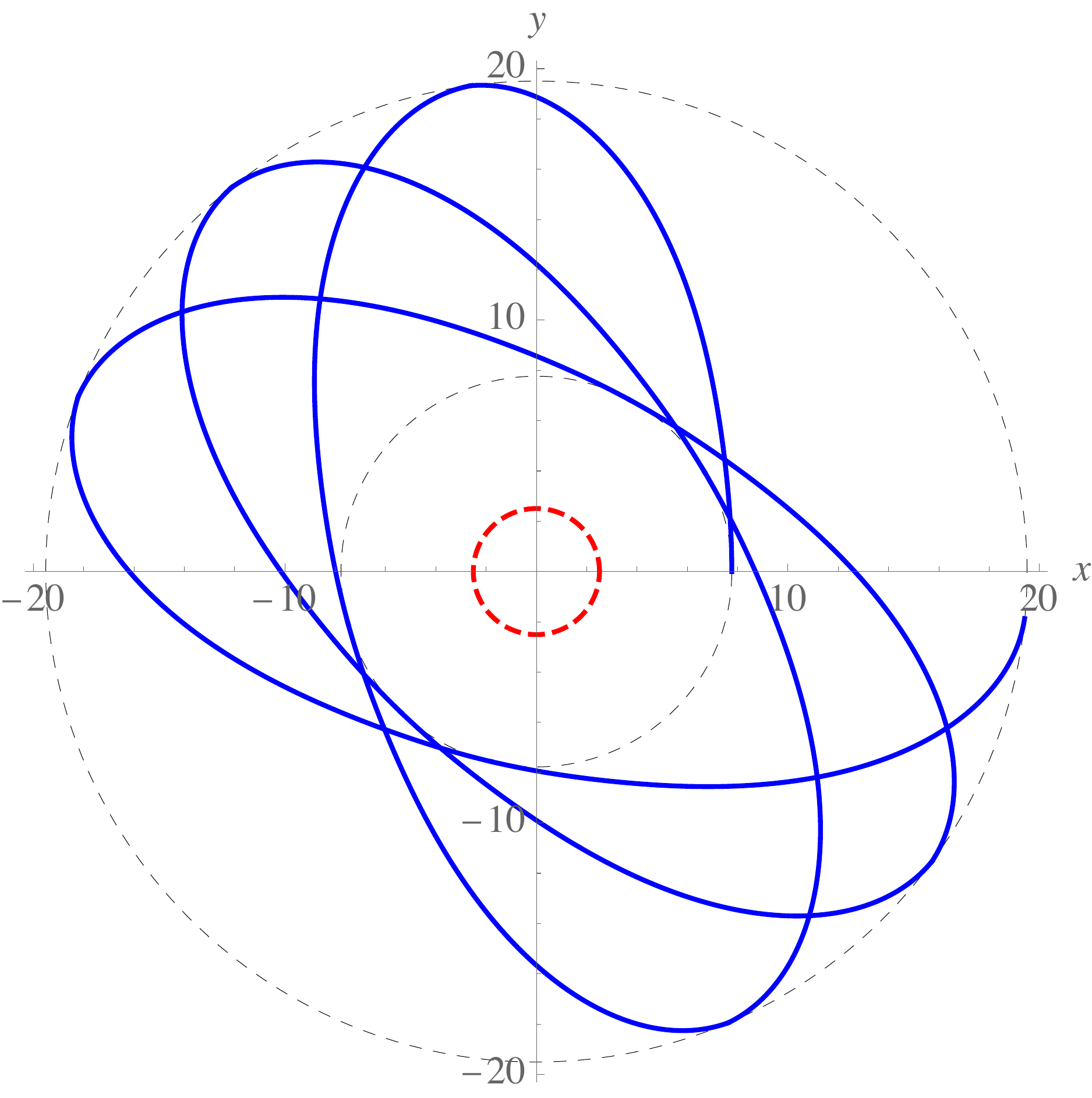}~(d)
		\includegraphics[width=4cm]{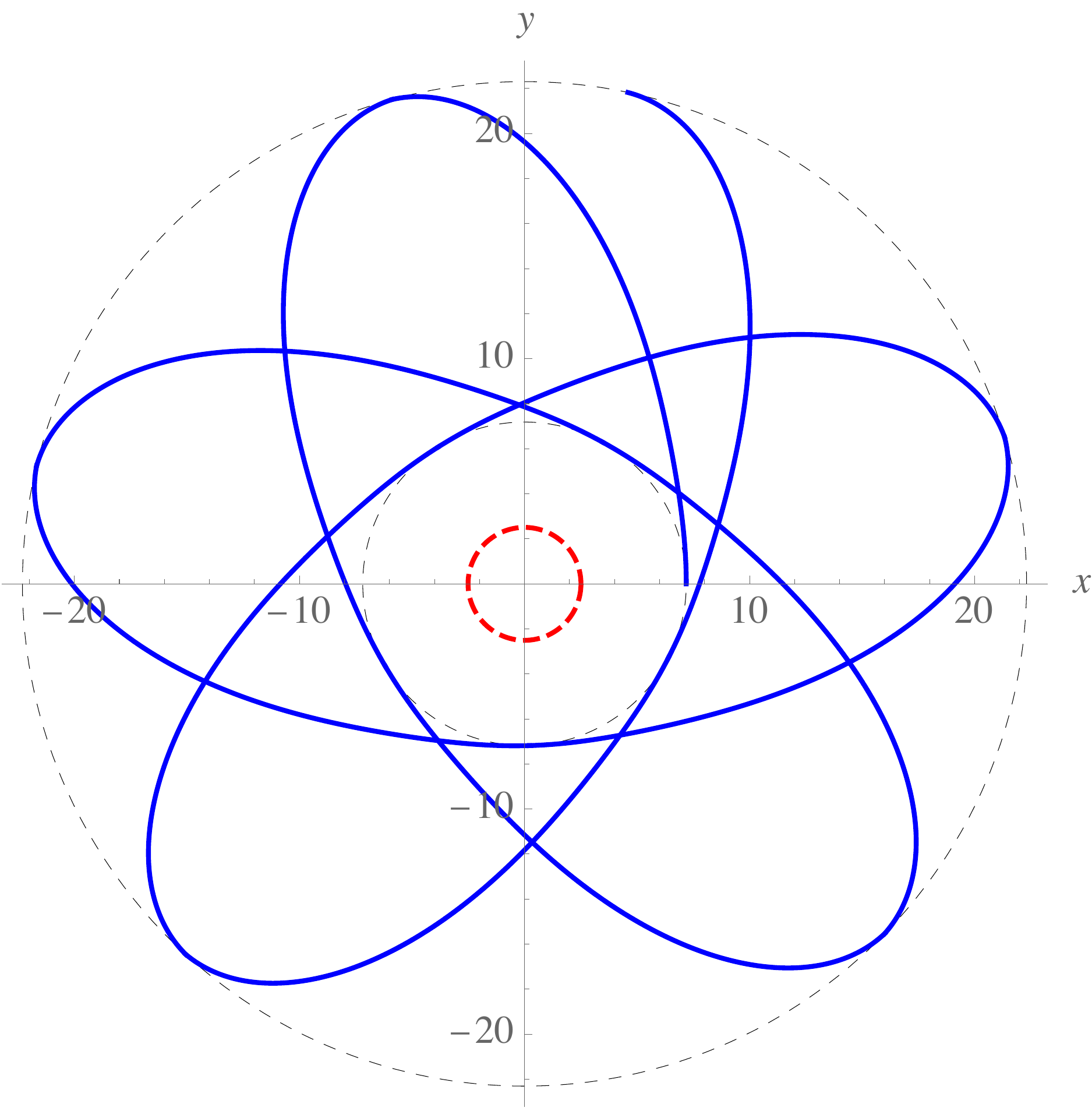}~(e)
		\includegraphics[width=4cm]{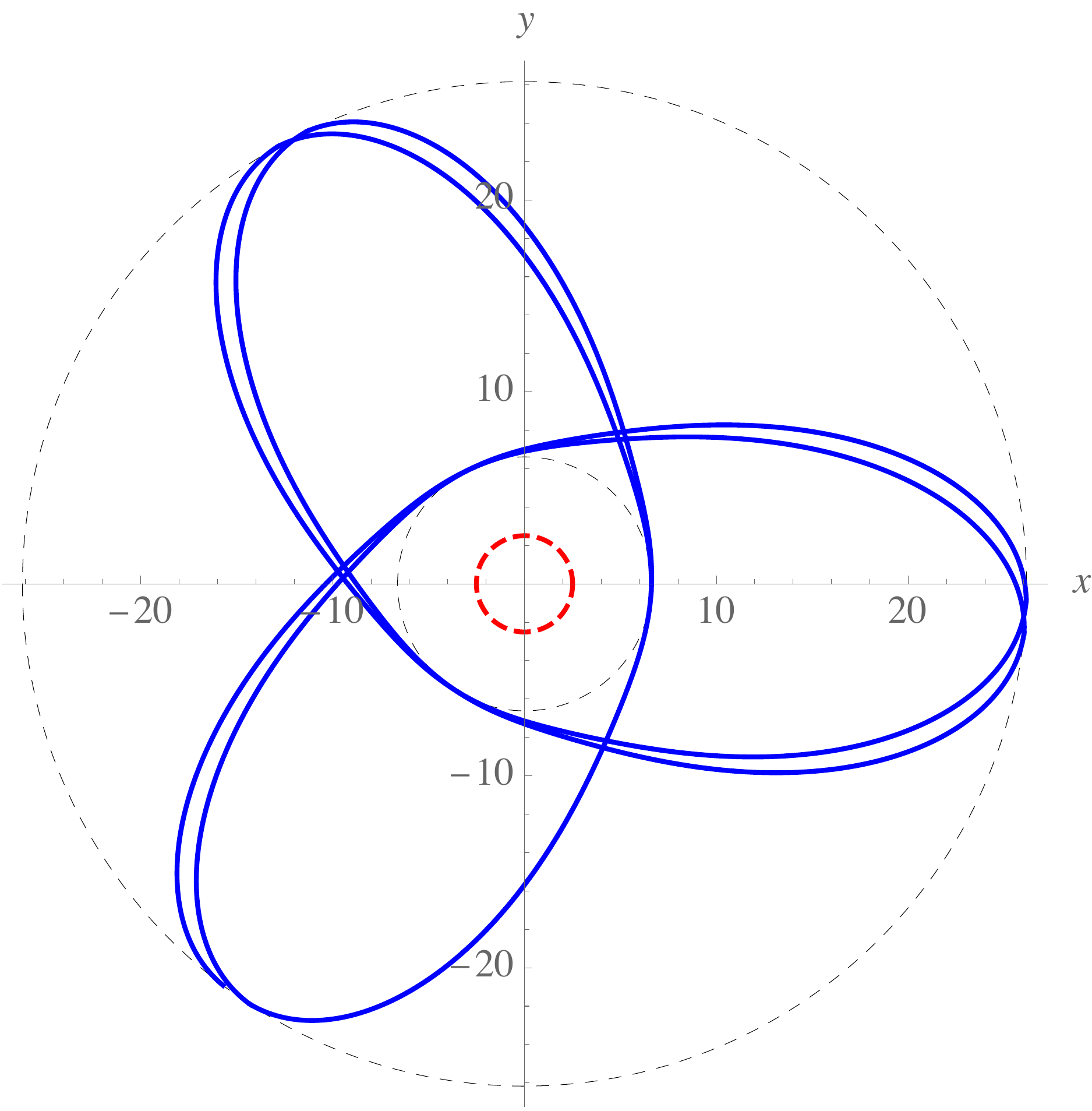}~(f)
		\includegraphics[width=4cm]{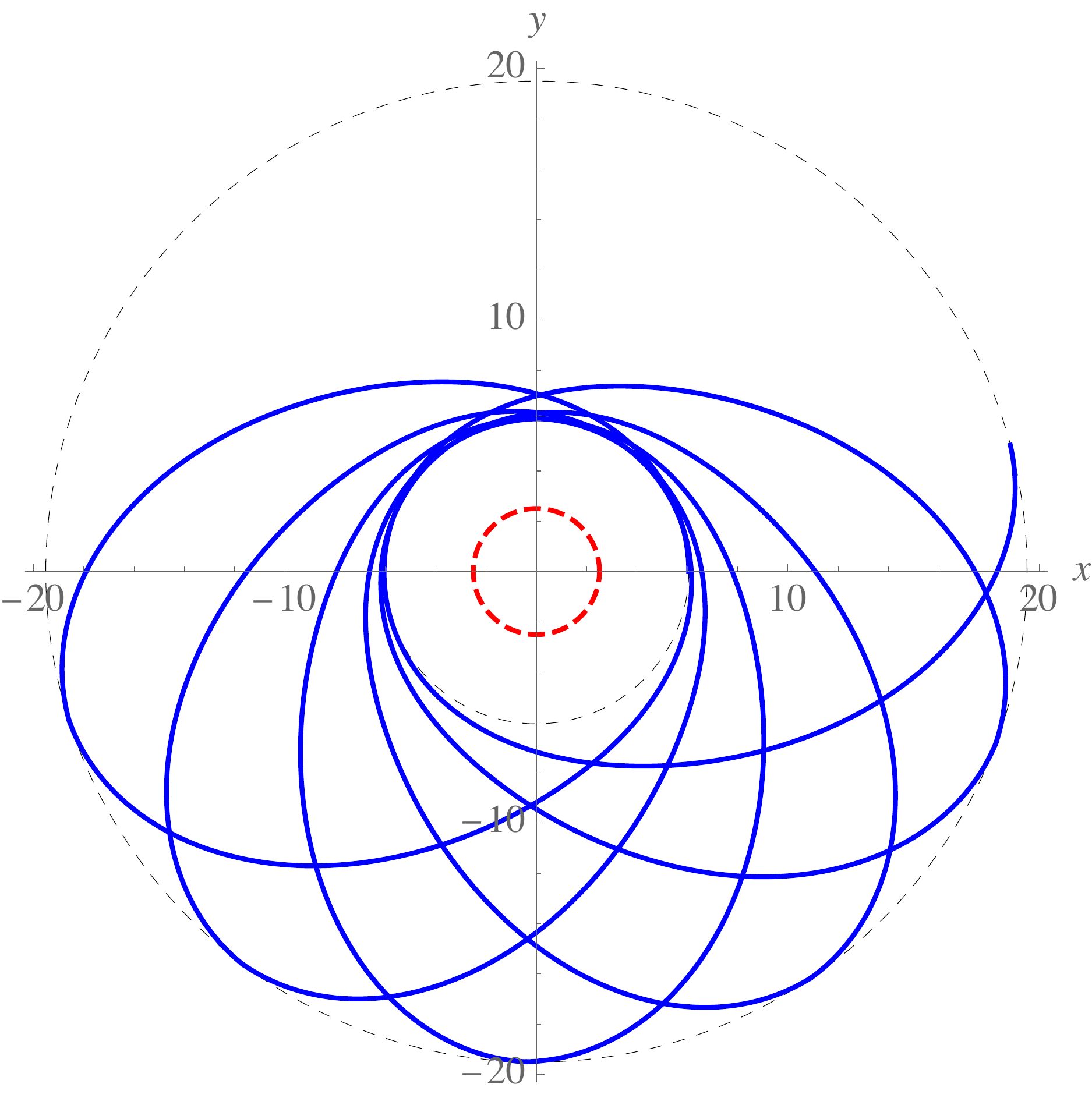}~(g)
	\end{center}
	\caption{Some examples of time-like planetary orbits, plotted in accordance with the effective potential in Fig.~\ref{fig:Vt-planetary}, for the energies (a) $E^2=0.709$, (b) $E^2=0.710$, (c) $E^2=0.712$, (d) $E^2=0.714$, (e) $E^2=0.716$, (f) $E^2=0.718$ and (g) $E^2=0.7194$. The inner and outer dashed circles indicate, respectively, $r_P$ and $r_A$ for each energy level and the small red circle is $r_+$.}
	\label{fig:planetary-timelike}
\end{figure}

For particles reaching at the turning point $r_A$ (or $r_P$) with the initial azimuth angle $\phi_0=0$, the planetary orbits are then characterized by the equation
\begin{equation}
\phi(r)= -L\int_{r_A}^r\frac{\ed r}{\sqrt{P_5(r)}},
    \label{eq:planetary_integral}
\end{equation}
which contains a hyper-elliptic integral on its right hand side. After applying specific intergation methods and manipulations, this integral results in the solution (see appendix \ref{app:B-nube})
\begin{equation}
\phi(r)=\frac{2L}{\sqrt{l^3 \gamma}}\sqrt{1-\frac{r}{r_A}}
F_D^{(4)}\left(\frac{1}{2},\frac{1}{2},\frac{1}{2},\frac{1}{2},\frac{1}{2};\frac{3}{2};c_1,c_2,c_3,1-\frac{r}{r_A}\right),
    \label{eq:planetary-solution}
\end{equation}
in which $l^3=(r_D-r_A)(r_A-r_P)(r_A-r_F)$, and $F_D^{(4)}$ is the incomplete Lauricella hypergeometric function of the fourth order, which is defined in the context of the integral equation \cite{Exton:1976,Akerblom:2005}
\begin{equation}
\int_0^{1-\frac{r}{r_A}}   u^{-\frac{1}{2}} (1-u)^{-\frac{1}{2}}\prod_{i=1}^{3}(1- x_i u)^{-b_i}\, \ed u = 2\sqrt{1-\frac{r}{r_A}} F_D^{(4)}\left(\frac{1}{2},b_1,b_2,b_3,\frac{1}{2};\frac{3}{2};c_1,c_2,c_3,1-\frac{r}{r_A}\right),
    \label{eq:FD(4)-rA}
\end{equation}
where $b_1=b_2=b_3=\frac{1}{2}$, and
\begin{subequations}
\begin{align}
& c_1=-{r_A\over (r_D-r_A)},\\
& c_2=  {r_A\over (r_A-r_P)}, \\
& c_3= {r_A\over (r_A-r_F)}.
\end{align}
\end{subequations}
In order to simulate the planetary orbits, we use the solution \eqref{eq:planetary-solution} to make a list of points $(\phi(r_t),r_t)$ in the domain $r_P<r_t<r_A$, and then we do the inversion by means of  numerical interpolations. In Fig. \ref{fig:planetary-timelike}, some examples of planetary orbits have been plotted for some different ranges of energy $E_S^2<E^2<E_C^2$.

Furthermore, the solution \eqref{eq:planetary-solution} allows for the determination of the precession of the periapsis in planetary orbits. This precession is given by $\Phi_{\mathrm{pl}}=2\phi_{AP}-2\pi$, where $\phi_{AP}$ is the azimuth angle swiped between $r_A$ and $r_P$ during the bound orbit. This way, one obtains
\begin{equation}
\Phi_{\mathrm{pl}}=\frac{4L}{\sqrt{l^3\gamma}}\sqrt{1-\frac{r_P}{r_A}}
F_D^{(4)}\left(\frac{1}{2},\frac{1}{2},\frac{1}{2},\frac{1}{2},\frac{1}{2};\frac{3}{2};c_1,c_2,c_3,1-\frac{r_P}{r_A}\right) -2\pi,
\end{equation}
as the exact expression for the precession in the planetary orbits.

\subsubsection{OFK and OSK}\label{subsubsec:OFKOSK-timelike}

\begin{figure}[t]
	\begin{center}
		\includegraphics[width=8cm]{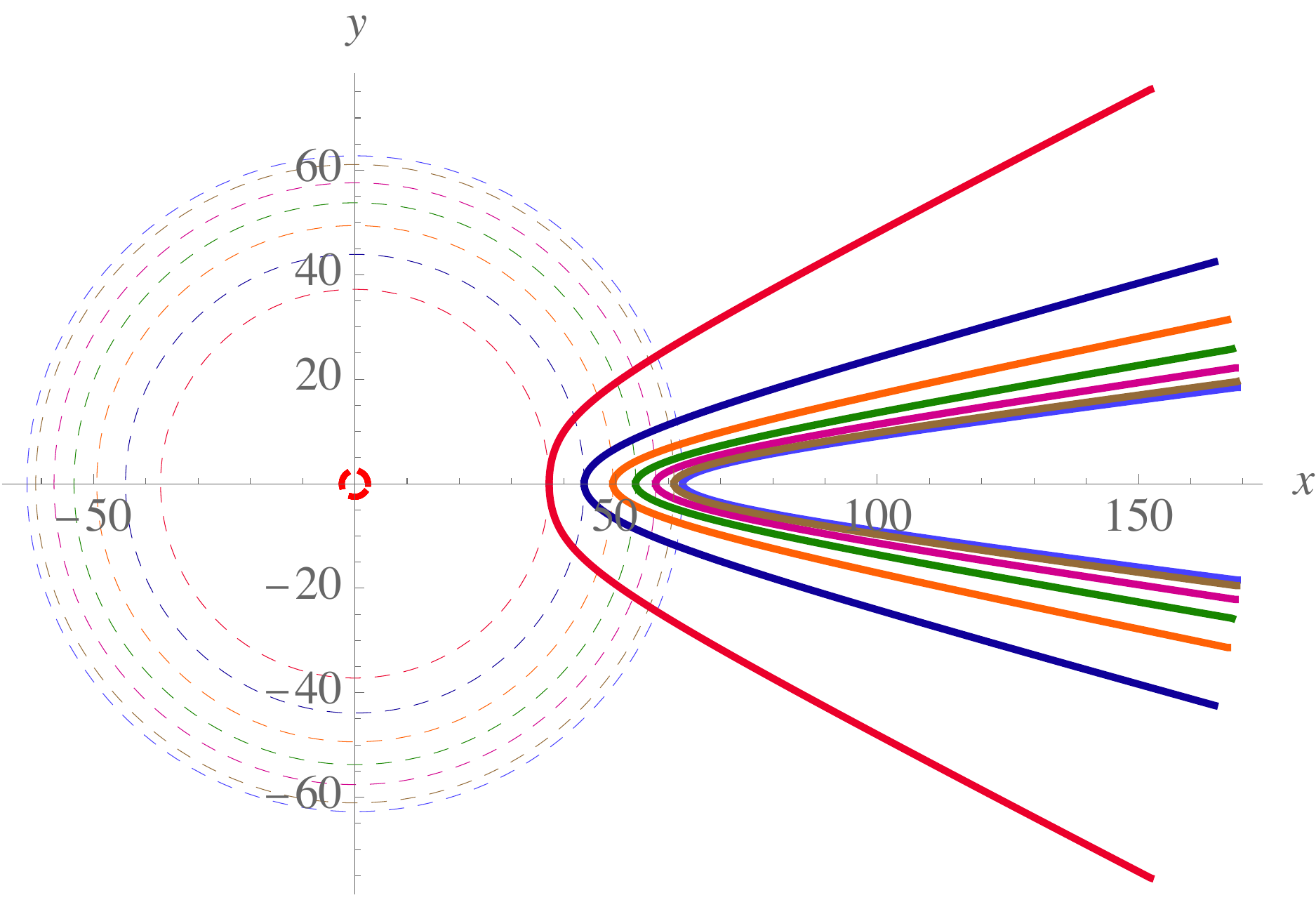}~(a)
		\includegraphics[width=8cm]{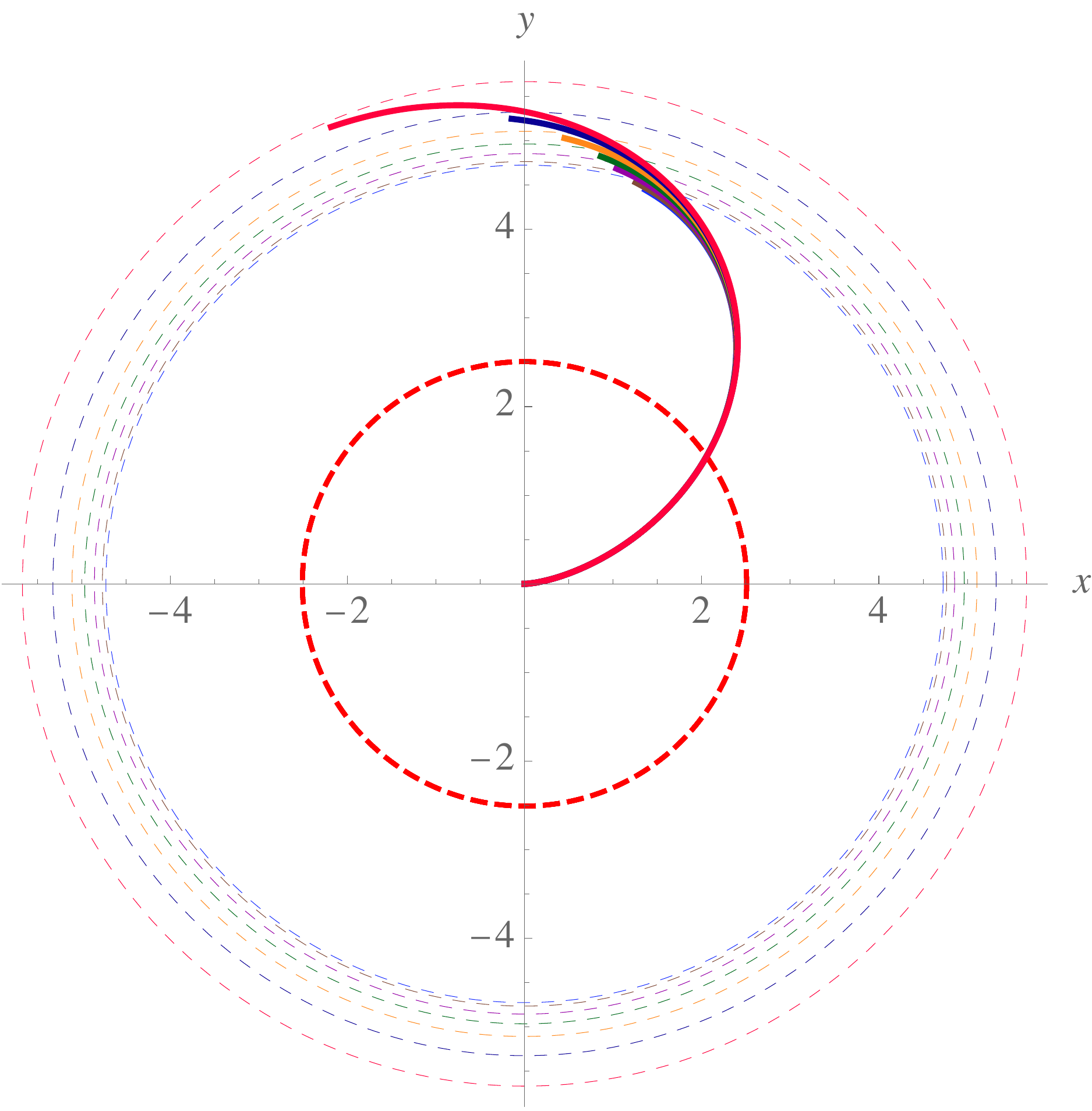}~(b)
	\end{center}
	\caption{The deflecting time-like trajectories, (a) OFK and (b) OSK plotted for the same values of energy as those used in Fig.~\ref{fig:planetary-timelike}. The thin dashed circles indicate $r_D$ for the OFK, and $r_F$ for the OSK. The value of $E^2$ decreases from the inner to the outer circles in the OFK, and from the outer to the inner circles in the OSK.}
	\label{fig:OFSK-timelike}
\end{figure}

The same effective potential in Fig.~\ref{fig:Vt-planetary}, offers the OFK and OSK for particles approaching at $r_D$ and $r_F$. So the same solution in Eq.~\eqref{eq:planetary-solution} holds for these kinds of orbits, and the only thing that changes is the point of approach. Therefore, following the same numerical method as applied for the planetary orbits, in Fig.~\ref{fig:OFSK-timelike}, the deflecting time-like trajectories have been plotted.

\subsubsection{Period of the circular orbits}\label{subsubsec:period-timelike}

\begin{figure}[t]
	\begin{center}
		\includegraphics[width=8cm]{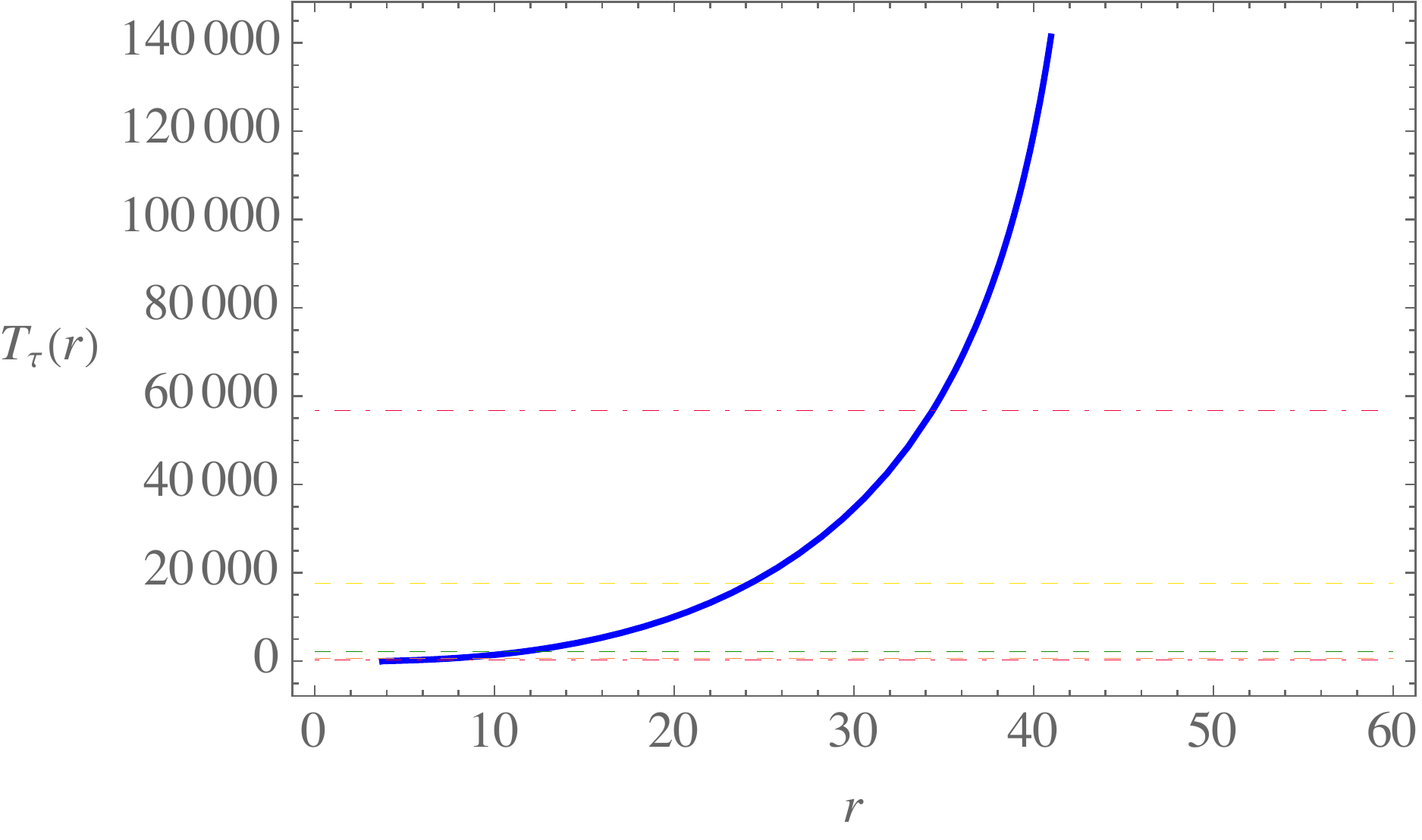}~(a)
		\includegraphics[width=8cm]{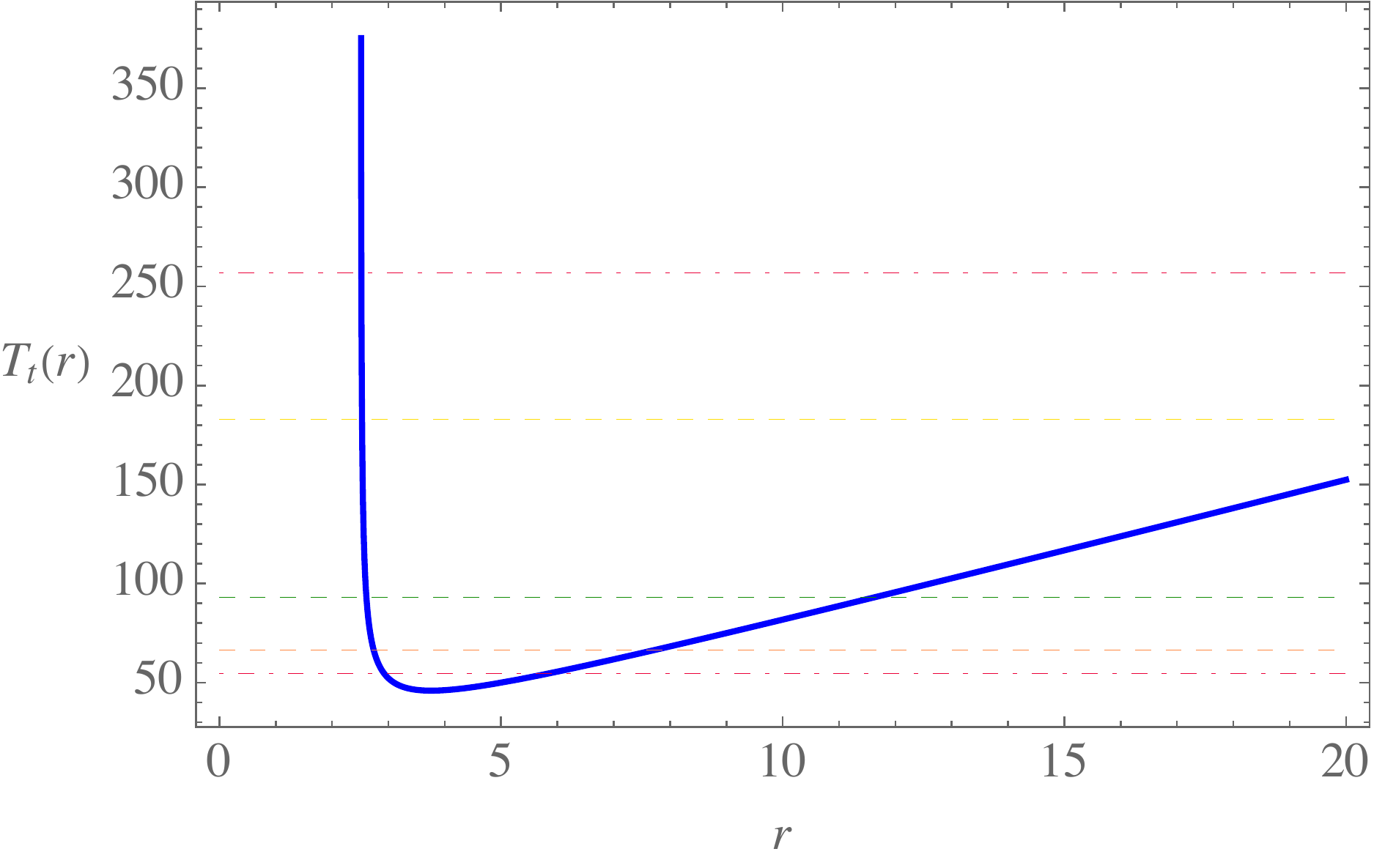}~(b)
	\end{center}
	\caption{The behavior of (a) the proper and (b) the coordinate periods of time-like circular orbits, in accordance with the values of $L$ as in Fig.~\ref{fig:Vt}. In the diagrams, the red dot-dashed lines correspond to unstable circular orbits at $r_{C_1}$ (lower line) and $r_{C_2}$ (upper line). The dashed lines correspond to the stable circular orbits at $r_\IS$ (orange), $r_\OS$ (yellow) and $r_S$ (green). }
	\label{fig:TtauTt}
\end{figure}

The period of stable and unstable circular orbits at the radial distance $r_{\mathrm{CO}}$, which can be stable or unstable, respectively at $r_S$ or $r_C$, can be calculated in the same way as in Subsect.~\ref{subsubsec:periodofrc}, and from Eqs.~\eqref{p1} and \eqref{p2}. In fact, by solving $V_t'(r)=0$ for $L$, one obtains the radial profiles 
\begin{eqnarray}
&& L(r) = r\sqrt{\frac{2M-\gamma r^2}{2(1-\alpha)r-6M-\gamma r^2}},\label{eq:LCO}\\
&& E^2(r) = \frac{2 \left[2 M-(1-\alpha -\gamma  r) r\right]^2}{r \left[2(1-\alpha)r-6M-\gamma r^2\right]},\label{eq:ECO}
\end{eqnarray}
for the circular orbits. Hence, by means of Eq.~\eqref{p1}, the proper period for the time-like circular orbits that exist at the distance $r_{\mathrm{CO}}$, becomes
\begin{equation}
T_{\tau}=2\pi r_{\mathrm{CO}}\sqrt{\frac{2(1-\alpha)r_{\mathrm{CO}}-6M-\gamma r_{\mathrm{CO}}^2}{2M-\gamma r_{\mathrm{CO}}^2}},
\label{eq:Ttau-timelike}
\end{equation}
to obtain which, we have let $L_{\mathrm{CO}}\equiv L(r_{\mathrm{CO}})$. Accordingly, at the Schwarzschild limit we have $T_{\tau}^{\mathrm{Sch}}=2\pi r_{\mathrm{CO}}\sqrt{\frac{r_{\mathrm{CO}}-3M}{M}}$. The period of the coordinate time, on the other hand, is obtained by using Eq.~\eqref{p2} at $r=r_{\mathrm{CO}}$, providing
\begin{equation}
T_t=2\pi r_{\mathrm{CO}}\sqrt{\frac{r_{\mathrm{CO}}}{r_{\mathrm{CO}}-2M-\alpha r_{\mathrm{CO}}-\gamma r_{\mathrm{CO}}^2}},
\label{eq:Tt-timelike}
\end{equation}
with the Schwarzschild limit $T_{t}^\mathrm{Sch}=2\pi r_{\mathrm{CO}}\sqrt{\frac{r_{\mathrm{CO}}}{r_\mathrm{CO}-M}}$. Taking into account the three important cases of $L=L_\IS, L_\OS$ and $L_C$ of the effective potential in Fig.~\ref{fig:Vt}, we have plotted the proper and coordinate periods in Fig.~\ref{fig:TtauTt}, together with their values at the distances $r_\IS$, $r_\OS$, $r_S$, and $r_{C_{1,2}}$.  

\begin{figure}[t]
	\begin{center}
		\includegraphics[width=8.5cm]{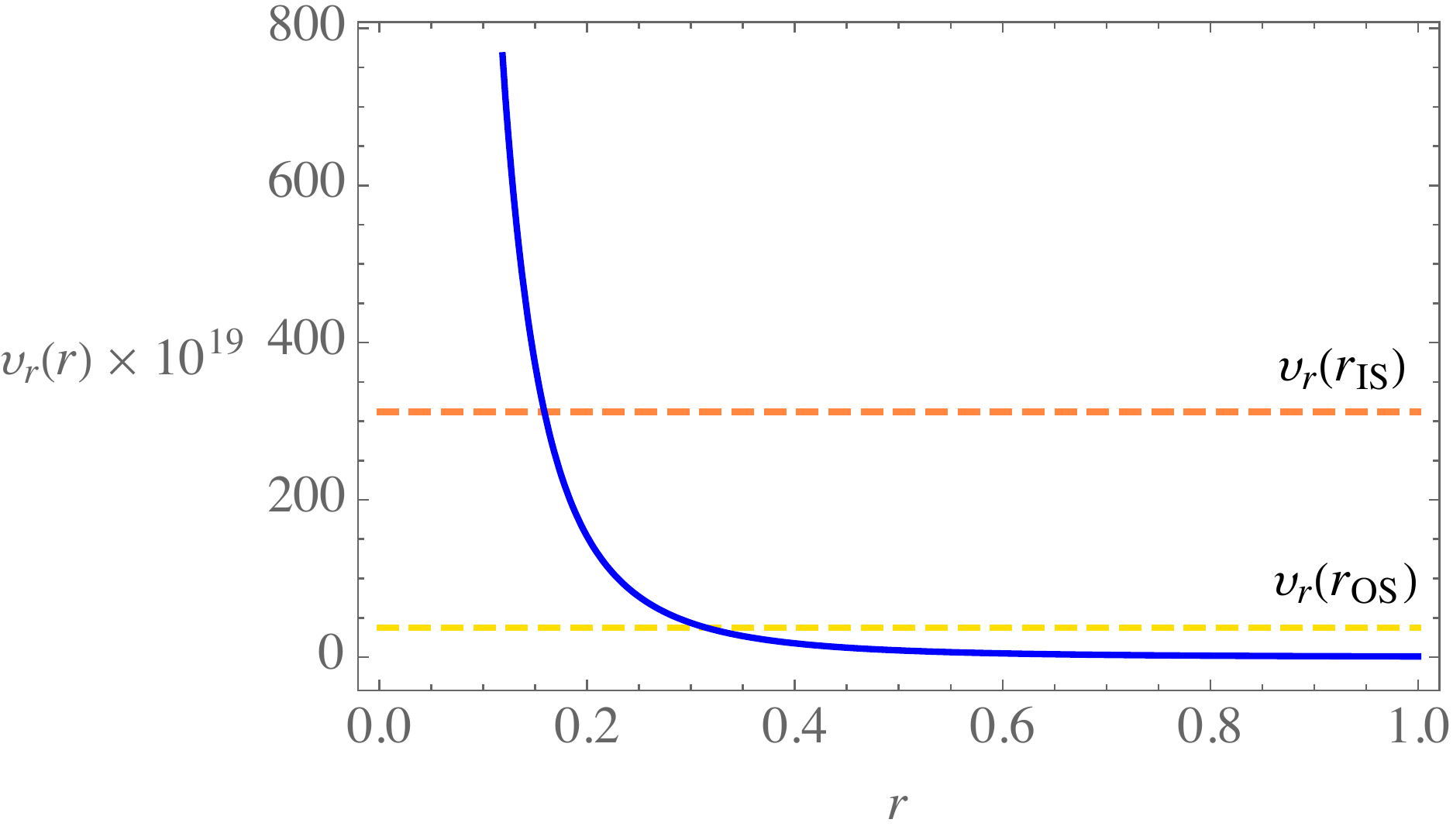}
	\end{center}
	\caption{The radial profile of the epicyclic frequency together with its values at the two radii of marginally stable orbits ($r_\IS$ and $r_\OS$), in accordance with the effective potential in Fig.~\ref{fig:Vt}.}
	\label{fig:Epicyclic}
\end{figure}

Note that, as they are marginally stable, orbits at the ISCO are sensitive to perturbations along the radial axis, in the sense that the orbiting particles may fall out of the stable orbits under certain circumstances. This way, one can define a limit, beyond which, the orbits become unstable. Such limit may be understood, indirectly, by mean of the radial \textit{epicyclic} frequency $\upsilon_r$, which is related to the formation of accretion disks around black holes (see Ref.~\cite{Abramowicz_foundations:2013} for a very good review). Therefore, $\upsilon_r$ is the frequency of the oscillations of the accreting particles along the radial direction, and is defined as \cite{Abramowicz_epicyclic:2005,Rayimbaev_dynamics;2021}
\begin{equation}
\upsilon_r^2=-\frac{1}{2g_{rr}}V_t''(r),
    \label{eq:EpicyclicFormula}
\end{equation}
given in terms of the effective potential. Taking into account the value of $L$ in Eq.~\eqref{eq:LCO} in the effective potential \eqref{eq:Vt}, we obtain
\begin{equation}\label{eq:upsilonr}
    \upsilon_r^2=\frac{\left[2M-(1-\alpha)r+\gamma r^2 \right]^2\left[12M^2-2M(1-\alpha)r-12M\gamma r^2+3\gamma (1-\alpha)r^3-\gamma r^4\right]}{r^3\left[6M-2(1-\alpha)r+\gamma r^2\right]^2}.
\end{equation}
In Fig.~\ref{fig:Epicyclic}, the radial profile of $\upsilon_r^2$ has been plotted together with its value at $r_\IS$.
Recently, the epicyclic frequency for quasi oscillations of massive test particles in circular accretions has been discussed in Ref. \cite{Mustafa_Osc:2021}, for the same black hole with an associated electric charge.
%
%

\subsubsection{Critical orbits}\label{subsubsec:critical-angular-timelike}

\begin{figure}[t]
	\begin{center}
		\includegraphics[width=8cm]{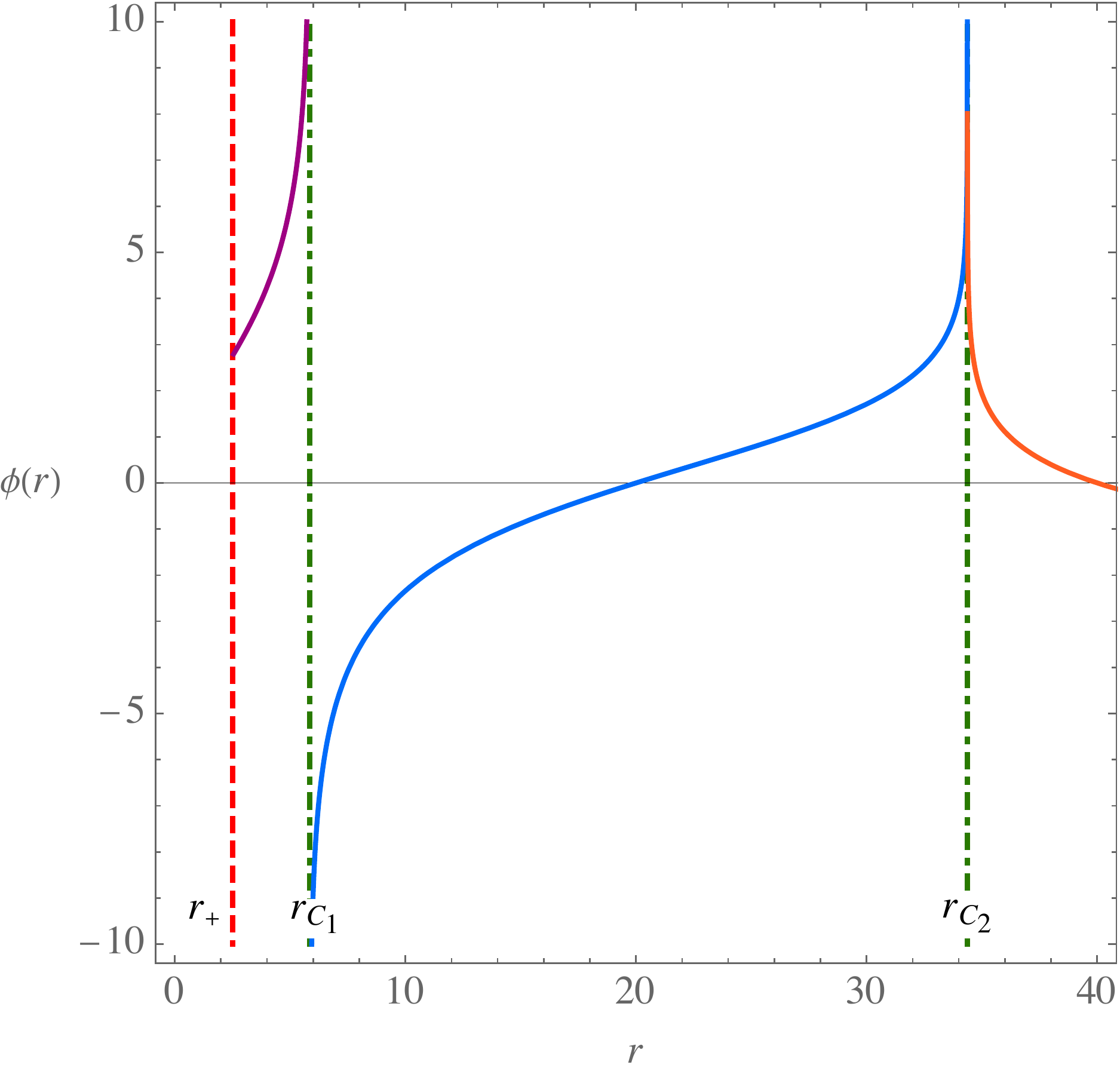}
	\end{center}
	\caption{The change in the $\phi$-coordinate during the time-like critical orbits, classified by colours, in accordance with different ranges of the initial radial distance $r_j$.}
	\label{fig:domainphi(r)}
\end{figure}

\begin{figure}[t]
	\begin{center}
		\includegraphics[width=8cm]{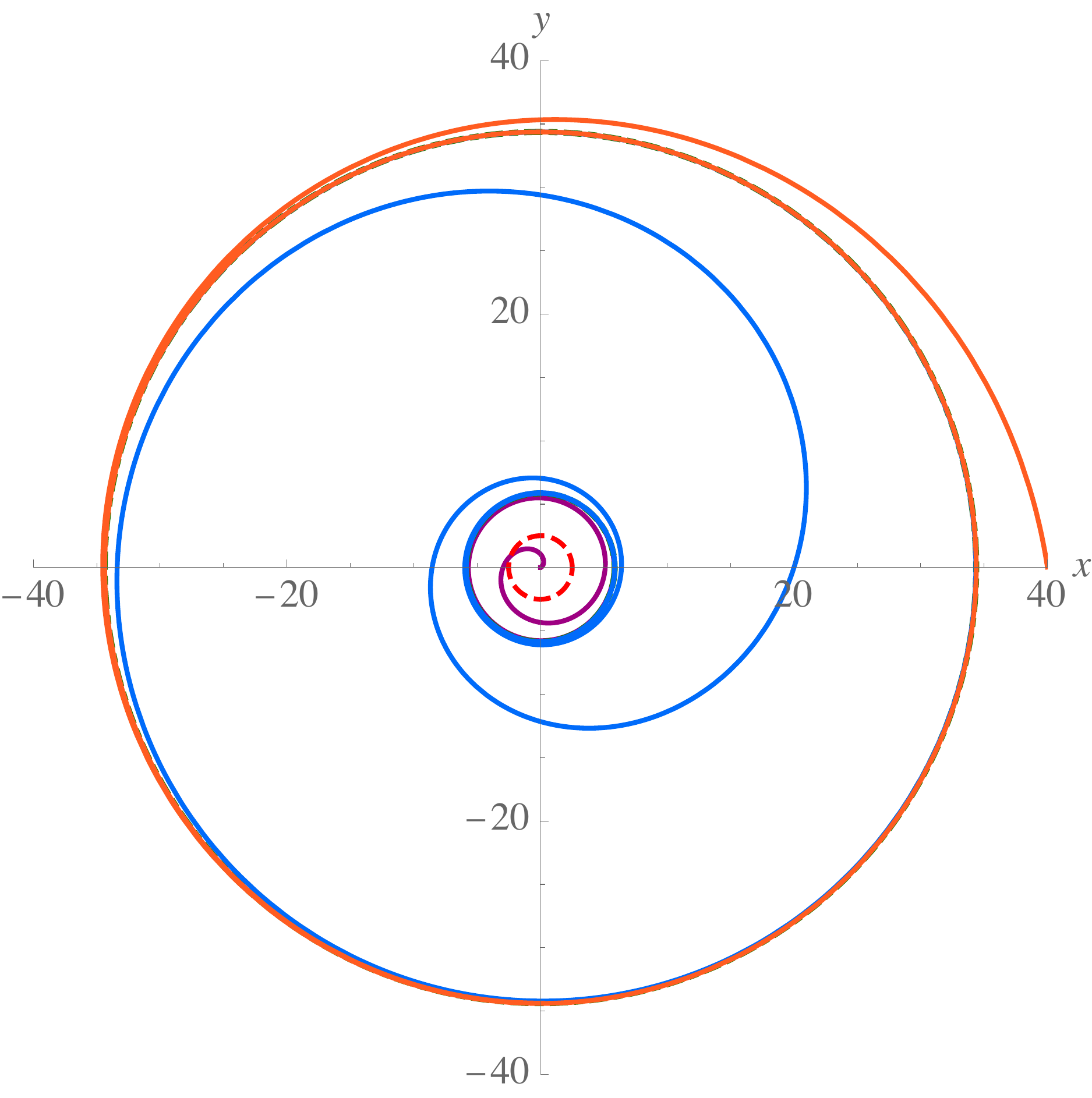}
	\end{center}
	\caption{The critical orbits of time-like geodesics for $L=L_C$, $E^2=E^2_C$, and different initial points $r_j$, in accordance with Fig.~\ref{fig:domainphi(r)}.}
	\label{fig:critical-timelike}
\end{figure}

Returning to the effective potential in Fig.~\ref{fig:Vt-planetary},  the two double roots $r_{C_1}$ and $r_{C_2}$ can provide a particular form of critical orbits. Once $E^2=E_C^2$, the characteristic polynomial can be recast as $P_5(r) = r(r-r_{C_1})^2(r-r_{C_2})^2$, for which, the angular equation of motion becomes
\begin{equation}
\phi(r)=L_C\int_{r_j}^r\frac{\ed r}{|r-r_{C_1}||r-r_{C_2}|\sqrt{r}},
    \label{eq:criticalEq0_timelike}
\end{equation}
for particles approaching from an initial point $r_j$. After some manipulations, it is found out that the critical orbits can be described in the context of the relation
\begin{equation}
Y(r)=\exp\left[\left(\frac{r_{C_2}-r_{C_1}}{2L_C}\right)\phi-\tilde{\varphi}_j\right],
    \label{eq:criticalEq1_timelike}
\end{equation}
in which
\begin{equation}
Y(r) = \left(\frac{1+\sqrt{\frac{r}{r_{C_2}}}}{1-\sqrt{\frac{r}{r_{C_2}}}}\right)^{\frac{1}{2\sqrt{r_{C_2}}}}
\left(\frac{1+\sqrt{\frac{r}{r_{C_1}}}}{1-\sqrt{\frac{r}{r_{C_1}}}}\right)^{-\frac{1}{2\sqrt{r_{C_1}}}},
    \label{eq:Y(r)}
\end{equation}
and
\begin{equation}
\tilde{\varphi}_j = \frac{1}{\sqrt{r_{C_2}}}\arctanh\left(\sqrt{\frac{r_j}{r_{C_2}}}\right)
-\frac{1}{\sqrt{r_{C_1}}}\arctanh\left(\sqrt{\frac{r_j}{r_{C_1}}}\right).
    \label{eq:tildevarphij}
\end{equation}
Therefore, the simulation of the orbits can be done by means of numerical interpolations. As before, the orbits can be classified in terms of the three cases $r_j>r_{C_2}$, $r_{C_1}<r_j<r_{C_2}$, and $r_+<r_j<r_{C_1}$. In Fig.~\ref{fig:domainphi(r)}, the domain of the changes in the $\phi$-coordinate during the critical orbits has been shown for each of these cases.
In Fig.~\ref{fig:critical-timelike}, the aforementioned three cases have been considered in Eq.~\eqref{eq:criticalEq1_timelike}, in order to simulate the possible forms of the critical orbits. 
As it is observed, the case of $r_j>r_{C_2}$ is indeed a COFK, whereas the orbits are completely confined between the two extremums when $r_{C_1}<r_j<r_{C_2}$. Finally, the case of $r_+<r_j<r_{C_1}$, is a COSK. Accordingly, distant observers will only be able to detect particles coming from $r_{C_2}$.

\section{Summary and concluding remarks}\label{sec:conclusions}


The study of particle dynamics is an interesting subject both from the theoretical and observational points of view. For example, as done in Ref.~\cite{Gravity:2020}, galactic astronomers can apply the formulations of periapsis precession for stars orbiting supermassive black holes, in order to compare the theoretical predictions with actual observations. In the same way, as far as the theoretical studies move forward, the observational evidences can be searched for, and vice versa. Hence, the theoretical predictions can help the applied scientists to obtain more insights on what they observe. In this paper, aiming at a theoretical study, we considered a Schwarzschild black hole with two associated cosmological terms stemming, respectively, in a quintessential field and a stringy cosmic fluid. Taking into account a special case for the EoS, namely $\omega_q=-\frac{2}{3}$, an extra linear gravitational potential is created. The resultant black hole admits an event and a cosmological horizon, that being null hypersurfaces, form the surfaces of infinite redshift and blueshift. This way, the radial profiles of the proper and coordinate times show two receding branches towards either of these horizons. Although the radial evolution of null and time-like geodesics for this black hole has been studied analytically in Ref.~\cite{Mustafa:2021}, here we presented an alternative approach to the analytical solutions reported in that paper, by employing strict elliptic integration methods that resulted in the \We{ian} expressions for the solutions. Furthermore, we categorized the radial orbits within the corresponding effective potential, and also in accordance with the turning points for which, we found explicit expressions. The radial profiles of the time axes were plotted for the particular cases of frontal scattering and critical motion. The analytical study of the angular trajectories for mass-less particles (photons) were first approached by means of the $\wp$-\We{ian} elliptic functions which enabled us analyzing the deflecting trajectories from the turning points which were also found explicitly. We continued the study of angular null geodesics by discussing the unstable circular (critical) orbits, which was the only remaining possible motion for mass-less particles. on the other hand, angular motion for massive particles appeared much more diverse in the sense that the effective potential could offer an ISCO and an OSCO, that confine the stable circular orbits. The corresponding angular momentums for these cases were obtained by taking into account the marginality condition $V''_t(r)=0$. There would be, therefore, some cases of two double roots for the characteristic polynomial. Accordingly, the choice of the angular momentum for the approaching test particles becomes of crucial importance in the analysis of the trajectories. To proceed, we adopted the switching value $L=L_C$, for which the two double extremums (maximums) of the effective potential are equal in the value. This case was analyzed in the context of several types of orbits that it could offer. For the case of planetary orbits, in addition to the periapsis and apoapsis, the characteristic polynomial has two other non-zero real roots that result in a hyper-elliptic integral for the azimuth angle. This integral was solved analytically in terms of the fourth order Lauricella hypergeometric function. To do the plots of the orbits, we obtained the inversion of this integral by means of numerical interpolations. The orbits show larger precession in the periapsis as the particle's energy approaches its critical value. The same method was used to plot the deflecting time-like trajectories. We also determined the proper and coordinate periods of stable circular orbits, and calculated the epicyclic frequency of accreting particles. We closed our discussion by analyzing the critical time-like orbits as they approach from three different initial points to either of the extremums. The simulations indicate that only the particles from the outer critical distance can escape to the infinity. In fact, the stability of circular orbits is an interesting subject of study for any black hole which is capable of the formation of accretion disks. Therefore, as a future study, this black hole can be investigated in more detail regarding the stability of accreting massive particles, as well as the lensing of its accretion disk and the possible higher order ring images.



\section*{Acknowledgements}
M. Fathi has been supported by the Agencia Nacional de Investigaci\'{o}n y Desarrollo (ANID) through DOCTORADO Grant No. 2019-21190382, and No. 2021-242210002. J.R. Villanueva was partially supported by the Centro de Astrof\'isica de Valpara\'iso (CAV).

\appendix

\section{Solving quartic equations}\label{app:A-nube}

The equation of the form
\begin{equation}
x^4+ax^3+bx^2+cx+d = 0,
    \label{eq:A.1}
\end{equation}
can be depressed by applying the change of variable $x\doteq z - \frac{a}{4}$, that yields
\begin{equation}
z^4+Az^2+Bz+C = 0,
    \label{eq:A.2}
\end{equation}
where
\begin{subequations}\label{eq:A.3}
\begin{align}
    & A = b-\frac{3a^2}{8},\\
    & B=c+\frac{a^3}{8}-\frac{ab}{2},\\
    & C=d+\frac{a^2b}{16}-\frac{3a^4}{256}-\frac{ac}{4}.
\end{align}
\end{subequations}
The method of finding the roots of Eq.~\eqref{eq:A.2}, has been given in the appendix C of Ref.~\cite{Fathi:2020sfw}.

\section{The angular solution for planetary orbits}\label{app:B-nube}

Applying the change of variable $x\doteq \frac{r}{r_A}$, the integral equation \eqref{eq:planetary_integral} changes its form to
\begin{equation}
\phi(x)=-\frac{L}{\sqrt{\gamma\,}}
\int_{1}^{x}\frac{\ed x}{\sqrt{P_5(x)}},
\label{eq:B.1}
\end{equation}
where
\begin{equation}
P_5(x)=x(r_D-r_Ax)(1-x)(r_Ax-r_P)(r_Ax-r_F).
\label{eq:B.2}
\end{equation}
Applying a second change of variable $x\doteq 1-u$, yields 
\begin{equation}
\phi(u)=\frac{L}{\sqrt{\gamma}}\int_{0}^{u}\frac{\ed u}{\sqrt{P_5(u)}},
\label{eq:B.3}
\end{equation}
in which
\begin{eqnarray}
P_5(u)&=&(1-u)\left[r_D-r_A(1-u)\right]u\left[r_A(1-u)-r_P\right]\left[r_A(1-u)-r_F\right]\nonumber\\
&=& u(1-u)(r_D-r_A)(r_A-r_P)(r_A-r_F)\left[1+\frac{r_A u}{r_D-r_A}\right]\left[1-\frac{r_A u}{r_A-r_P}\right]\left[1-\frac{r_A u}{r_A-r_F}\right]\nonumber\\
&\equiv& l^3 u(1-u)(1-c_1u)(1-c_2u)(1-c_3u).
\label{eq:B.4}
\end{eqnarray}
Hence, one can recast Eq.~\eqref{eq:B.3} as
\begin{equation}
\phi(u) = \frac{L}{\sqrt{l^3\gamma}}\int_0^{u} u^{-\frac{1}{2}}(1-u)^{-\frac{1}{2}}(1-c_1 u)^{-\frac{1}{2}}(1-c_2 u)^{-\frac{1}{2}}(1-c_3 u)^{-\frac{1}{2}}\, \ed u.
    \label{eq:B.5}
\end{equation}
The incomplete Lauricella function of order $n+1$, is defined in terms of the integral equation \cite{Akerblom:2005}
\begin{equation}
\int_0^{z} u^{a-1} (1-u)^{c-a-1}\prod_{i=1}^n (1-x_i u)^{-b_i}\, \ed u = \frac{z^{a}}{a}F_D^{(n+1)}\left(a,b_1,\dots,b_n,1+a-c;a+1;x_1,\dots,x_n,z\right).
    \label{eq:B.6}
\end{equation}
By doing a comparison between Eqs.~\eqref{eq:B.5} and \eqref{eq:B.6}, it is inferred that $a=\frac{1}{2}$, $c=1$ and $n=3$. Accordingly, we get $b_1=b_2=b_3=\frac{1}{2}$, and $x_1=c_1$, $x_2=c_2$ and $x_3=c_3$. This way, the solution in Eq.~\eqref{eq:planetary-solution} is obtained.

\bibliographystyle{ieeetr}
\bibliography{biblio_v1.bib}

\end{document}